\documentstyle{article}
\newcommand{\f}{\begin{equation}}
\newcommand{\ff}{\end{equation}}

\newcommand{\dthree}{d^{3}x}                   

\newcommand{\reals}{{\rm I\! R}}
\newcommand{\complex}{{\rm I \!\!\! C}}

\def\ut#1{\rlap{\lower1ex\hbox{$\sim$}}#1{}}        

\newcommand{\be}{\begin{equation}}
\newcommand{\ee}{\end{equation}}
\newcommand{\ben}{\begin{enumerate}}
\newcommand{\een}{\end{enumerate}}
\newcommand{\bcen}{\begin{center}}
\newcommand{\ecen}{\end{center}}

\newcommand{\bea}{\begin{eqnarray}}
\newcommand{\eea}{\end{eqnarray}}

\newcommand{\ie}{{\it i.e.\mbox{$\;$}}}

\newcommand{\etal}{{\it et.al.\mbox{$\;$}}}

\begin{document}

\rightline{\large DRAFT}
\vfil
\centerline{\Large \bf Fermions in Quantum Gravity}
\vfil
\centerline{\normalsize\it Hugo A. Morales-T\'ecotl${}^1$ and Carlo
Rovelli${}^2$}

\centerline{${}^1$\ \it SISSA, Strada Costiera 11, Trieste, Italy}

\centerline{${}^2$\ \it   Department of Physics, University of
Pittsburgh, Pittsburgh,15260 U.S.A.}
\centerline{${}^2$\ \it   Dipartimento di Fisica, Universita' di Trento. INFN,
Sezione di Padova, Italia.}

\date{today}

\vfil

\small
\centerline{\bf Abstract}
\vskip.5cm

We study the quantum fermions+gravity system, that is, the
gravitational counterpart of QED.   We start from the standard
Einstein-Weyl theory, reformulated in terms of Ashtekar variables;
and we construct its non-perturbative quantum theory by
extending the loop representation of general relativity.

To this aim, we construct the fermion equivalent to the loop
variables, and we define the quantum theory as a representation of
their Poisson algebra. Not surprisingly, fermions can be
incorporated in the loop representation simply by including open
curves into ''Loop space'', as expected from lattice
Yang-Mills theory. We explicitely construct the diffeomorphism and
hamiltonain operators.
The first can be fully solved as in pure gravity. The second is
constructed by using a background-independent regularization
technique.  The theory retains the clean geometrical features of the
pure quantum gravity.  In particular, the hamiltonian constraint
admits the same simple geometrical interpretation as its pure
gravity counterpart: it is the operator that shifts curves along
themselves (''shift operator''). Quite surprisingly, we believe, this
simple action codes the full dynamics of the interacting
fermion-gravity theory.

To unravel the dynamics of the theory we study the evolution of the
fermion-gravity system in the physical-time defined by an
additional coupled (''clock''-) scalar field. We explicitely construct
the Hamiltonian operator that evolves the system in this physical
time. We show that this Hamiltonian is finite, diffeomorphism
invariant,
and has a simple geometrical action confined to the intersections
and the end points of the ''loops''. The quantum theory of
fermions+gravity evolving in the clock time is finally given by the
combinatorial and geometrical action of this Hamiltonian on a set of
graphs with a finite number of end points. This geometrical action defines  the
"topological Feynman rules" of the theory.

\vfil
\normalsize
\clearpage
\pagenumbering{arabic}

\section{Towards Quantum Gravitational Dynamics}

{\it 1.1 Motivations}
\vskip .5cm\noindent

The theoretical description of the Quantum Electromagnetic Field
was developed in the thirties, shortly after the birth of quantum
theory.  In spite of some immediate successes as the
justification of the existence of photons, it is not untill the
construction of the full quantum theory of the fermions
+ electromagnetism system, nameley QED, that the full power of this
theory became clear.     Recently, a certain number of advances
toward the construction of Quantum General Relativity (GR) have
been achieved
\cite{carlolee,weave,gravitons,carloobserves,RSphysham} (for an
introduction see \cite{review,abhayreview,review-ls}).  However,
we suspect that this theory too will
express its full physical value only when a fully interacting matter
+ gravity theory is constructed, and, in particular, when the
realistic fermions+gravity theory is constructed.   By analogy with
QED, we shall denote the quantum theory of fermions + gravity as
Quantum Gravitational Dynamics,  or QGD.

There is a number of reasons for suspecting that matter couplings
are needed for clarifying the Quantum Gravity puzzle.  The first of
these, is that it is very difficult to write fully gauge
invariant quantities on the phase space of General Relativity alone,
due to
diffeomorphism invariance \cite{carlolee,observable}.  Equivalently,
it is extremely difficult
to imagine well-defined experiments to be performed on the
gravitational field {\it alone. }   Thus, in the pure gravity case we
are in the funny situation of constructing a theory, but not being
sure of what precisely should we ask to the theory -- a situation
quite familiar,
we believe, to anybody working in non-perturbative quantum
gravity.   On the other side, diffeomorphism invariant quantities, as
well as realistic experiments described by those quantities, can be
constructed in a relative simple fashion in the presence of
dynamically coupled matter \cite{observable,qobservable,tradition}.
For instance, the Solar
System, or a binary pulsar emitting gravitational radiation, are
examples of matter + gravity systems that we understand well as
far as measurements are concerned: we know exactly which
meaningfull observables we can measure.  If these
observables could be measured with Planck-scale sensitivity, then
these systems could be seen as quantum gravity laboratories.
Clearly, we need a matter + gravity quantum theory in order to
describe them theoretically.

A second reason for coupling matter to quantum gravity, is given by
the peculiarity of the non-perturbative quantum theory of gravity.
In the Loop Representation \cite{carlolee,gambini}, the theory has a
characteristic
geometrical structure:  Quantum states of gravity are classified by
Knot Theory \cite{carlolee}, and the dynamics can be represented in
a fully
combinatorial-algebraic fashion on Knot Space \cite{RSphysham}.
These features are
not accidental, but rather are consequences of diffeomorphism
invariance; equivalently, they are related to the fact that the Loop
Representation is a genuine background independent quantum field
theory.   Since the theory relies on these geometrical structure, it is
mandatory to check whether these structures are lost when further
fields are coupled.  If so, doubts could be cast on the value of the
Loop Representation of Quantum GR.

\vskip .5cm\noindent
{\it 1.2 Introducing fermions in the loop representation}

\vskip .3cm\noindent
Motivated by the considerations above, we have studied the
fermions + gravity system, or QGD, in the Loop Representation.  The
choice of fermions is motivated by realism, but also by the
fact that they are very natural objects in the Ashtekar formalism.
The study of the gravitational interaction of fermions in the light of
General Relativity, goes back to Dirac and Sciama
\cite{diracsciama}, and has more
recently been investigated by Nelson and Teitelboim \cite{nelson}
using
a canonical approach.  The theory has been formulated in terms of
Ashtekar variables in ref. \cite{abhayfermions};  we review this
formalism
below.  We then construct the quantum theory following the lines
along which the quantum theory of pure General Relativity was
constructed.   We
define the natural extension of the loops observables to fermions
(these are given by a parallel transport operator associated to an
open curve contacted with fermions sited at the end points of the
curve), study their Poisson algebra, and define the quantum theory as
a linear representation of this algebra.  In analogy with the pure
gravity case, we show that the resulting representation can also be
heuristically obtained from a naive Shr\"odinger-like representation
by means of a (ill defined) Loop Transform.

The loop representation of QGD turns out to be a very natural
extension of the pure gravity case, obtained by including {\it open
curves\ } into Loop Space.   This is certainly not surprising, since
the kinematics of the loop representation can be seen as the
continuum version of the Wilson-Kogut construction in lattice
Yang-Mills theory \cite{wilson}, and from the work of Gambini and
collaborators \cite{gambini} where fermions are represented by
the end points of
open lines of flux on the lattice.   In the rest of the paper, we will
denote both open and closed curves as {\it loops\ }, disregarding
consistency with the dictionary.   It is not difficult to solve the
diffeomorphism constraint on the resulting state space.  The
complete
classification of the solutions is given by a generalization of the
Knot Classes of the pure gravity case -- the new classes include
graphs with an arbitrary number of intersections and open ends.
Thus, quantum states of QGD admit the same kind of topological
description as the states of pure Quantum GR, contrary to the fear
that this aspect of the Loop Representation could be lost in presence
of matter couplings.  We view this as an encouraging result, though
a result that could have been anticipated.

On the other side, the results we obtain about the {\it dynamics\ } of
the theory are unexpected and, we believe, rather surprising.  The
dynamics is given by the hamiltonian constraint, which includes
the fermion kinetic energy and the fermion-gravity interaction.  We
construct in this paper the corresponding
quantum operator, and its action on
the loops turns out to have an extremely simple geometrical
interpretation:  The hamilonian constraint operator essentially
"shifts"  loops along their tangent.  This same simple geometrical
action of the hamiltonian constraint was recognized in the context
of pure gravity in ref.\cite{carlolee}.    The surprising result here is
that the very same action, extended to {\it open\ } loops, codes the
kinetic fermion energy and the fermions-gravity interaction. This
was first noticed by one of us in ref.\cite{hugo}.

\vskip .5cm\noindent
{\it 1.3 Introducing a clock}

\vskip .3cm\noindent
While suggestive, the above construction is not fully satisfactory
for three reasons. First there is a divergence in the action of the
hamiltonian operator which is difficult to control. Second, in spite
of the simplicity of the action of the hamiltonian operator, we have
not been able to solve the corresponding quantum constraint
equation.  Third, the presence of the fermions does not takes us
completely out from the difficulties of constructing gauge invariant
observables: it simplifies the task of finding three-dimensional
diffeomorphism invariant quantities, but it does not help with the
problem of finding quantities that commute with the hamiltonain
constraint. Thus, the actual physical content of the theory is still
quite unaccessible, as it is in pure gravity.

To face these problems, we take one further step. We combine
our results on the fermions with the results obtained in
ref.\cite{RSphysham}.
In that paper, the idea was proposed to unravel the dynamics of
quantum gravity by coupling a scalar field, which could behave as a
clock-field, following a long tradition \cite{tradition} of ideas of
using matter for simplifying the gravitational theory analysis (see
\cite{observable,qobservable,time,hypotesis,model,clock}).    It is
shown in Ref. \cite{RSphysham}
that by a suitable gauge fixing one can
express the dynamics of gravity as intrinsic evolution with respect
to the intrinsic time (or physical time) defined by the scalar field.
This evolution is explicitely generated on the state space by a
hamitonian operator $\hat H$.  Here, we
extend this construction to fermions.  Namely, we consider the
generally covariant gravity+fermions+scalar field system, we solve
with respect to the scalar field, so that the hamiltonian constraint
disappears from the theory, and is replaced by a genuine
diffeomorphism invariant hamiltonian that evolves both gravity and
fermions in the scalar field clock time.    We explicitely construct
the {\it quantum hamiltonian\ }  operator  $\hat H$  (as opposed to
hamiltonian constraint operator), by making use of the
regularization techniques on manifolds that have recently been
introduced \cite{weave} in quantum gravity.  In the fermion sector,
we recover
in this way the simple action described above, namely the shift of
the loops along themselves.  However, now the resulting operator is
fully {\it diffeomorphism invariant}, and {\it finite\ }.

\vskip .5cm\noindent
{\it 1.4 QGD}

\vskip .3cm\noindent
The resulting QGD is then given by a set of quantum states
represented by graphs with a finite number of intersections and open
ends, and by an Hamiltonian operator that acts in a simple
geometrical and combinatorial fashion on these graphs. Matrix
elements
of this Hamiltonian can be interpreted as {\it first order transition
amplitudes between the graph states in a time dependent
perturbation expansion in the clock time.\ } The explicit computation
is complicated by the need of extracting the square root of an
infinite matrix, a task we expect could be solved order by order.  In
the present paper, we only begin the explicit computation of matrix
elements of the operator.

The picture of Quantum Gravitational Dynamics that begins to
emerge from this construction has a simple and perhaps appealing
general structure:    A graph with two open ends, say, represents two
fermions interacting gravitationally among themselves, and with the
surrounding gravitational field.   With the machinery developed in
this paper we could (at least in principle) follow the quantum
evolution of this system in clock time.

The paper is organized as follows.  In section 2, we review the
classical Einstein-Weyl theory in Ashtekar form, and we
introduce the loop variables and their extension to fermions.  In
section 3, we define the quantum theory, and we discuss
and solve the quantum diffeomorphism constraint. In section 4, we
give a preliminary discussion of the Hamiltonian quantum constraint
and its surprising simplicity -- the fully consistent construction
needs the clock-field.  In section 5, we recall the main ideas of the
clock-field construction, we define the gravity+fermions+clock-
field theory, we study the regularization of the
Hamiltonian, and we obtain the rigorous form of the Hamiltonian
operator.   In section 6, we describe resulting structure of QGD, we
collect a certain number of comments on the general construction,
we discuss the lines along which the theory should be further
developed, and we summarize our results.

The signature of the spacetime metric is $(-+++)$. Throughout units
are used in which $G=c=\hbar=1$, except in the last section.

\section{Einstein-Weyl theory in Ashtekar form}

We consider the system formed by general relativity and a Weyl
fermion field.  We describe general relativity in terms of Ashtekar
variables \cite{ashtekar}; and the fermion in terms of a two-
component, massless spinor field  ${\psi}^{A}(x), A=1,2$.

A word should be spent concerning the coupling between the
fermions and gravity. Matter spinor fields coupled to gravity in the
Ashtekar's formalism  have been studied in refs.
\cite{abhayfermions}  (see also  \cite{Esposito-Morales-Kim}).
Ashtekar  \etal~  have added a quartic terms to the  minimally
coupled fermion-gravity action in order to get the equations of
motion of the Einstein-Weil torsion-free theory.  As made clear in
\cite{Esposito-Morales}, the minimally coupled action {\it without}
quartic terms, yelds the Einstein-Cartan theory, in which fermions
act as a source of (non-propagating) torsion.  \footnote{In
the theory with the quartic interaction one can naturally define a
connection which is the sum of the torsion-free selfdual connection,
and a matter generated term; the equations of motion are then
structurally equal to the Einstein-Weil equations, but with the
selfdual connection replaced by the full connection. See
\cite{Esposito-Morales}  for more details.}. For simplicty, we
consider here the minimally coupled theory. The system represents
for instance neutrinos in a dynamical
spacetime.  We expect that the extension to Dirac fermions, and to
the theory with the quartic interaction should be straight forward.

In recent years it has become conventional to work with (classical)
spinor fields which are Grassmann valued.  We (reluctantly) decided
to follow this practice because it simplifies the bookkeeping of the
signs in promoting Poisson brackets to (anti-) commutators (since
we do not use canonically conjugate variables as basics observables
for the quantization). Thus, we assume the fields ${\psi}^{A}(x)$
anticommute.

Spacetime is taken as a manifold $M$ with the topology
$\Sigma\times \reals$, with $\Sigma$ compact.  We indicate
spacetime indices (four dimensional as well as three dimensional)
with lower case latin letters. Over and under tildes denote densities
of weight $+1$ or \ $-1$ respectively (on $M$ as  well as on
$\Sigma$).  Since we are dealing with spinors, it is natural to adopt
the spinor version of the Ashtekar formalism.
The gravitational variables are then the spacetime $SL(2,\complex)$
soldering form ${}^{4}\sigma^{a~A'}_{~A}(x)$, that is the spinor form
of the tetrad field, and the $SL(2,\complex)$ Ashtekar connection
\cite{ashtekar}, which has only  unprimed indices,
${}^{4}\!A_{aA}^{~~B}(x)$, and which is interpreted as the (internal-)
selfdual part of the spin connection.
We give the explicit relation with the vector quantities below.
The Lagrangian of the system is \cite{abhayfermions,action,tedj}
\be
{\cal L}_{\rm Einstein-Weyl} = ({}^{4}\sigma)
{}^{4}\sigma^{a~A'}_{~A}
                                {}^{4}\sigma^{b}_{~BA'} {}^{4}\!F_{ab}^{~~AB}
+ \sqrt{2} \, ({}^{4}\sigma) {}^{4}\sigma^{a}_{AA'}
  \bar{\psi}^{A'} {}^{4}{\cal D}_{a}\psi^{A}.
\ee

In order to construct the quantum theory, we set the theory in
canonical form.  This can be done in a fully covariant way by taking
the space of the solutions of the equations of motion as the phase
space, and defining the appropriate symplectic struture over it.
However, we follow here the equivalent, but simpler and more
conventional, approach, based on the  3+1 decomposition, even if it
obscures the natural general covariance of the hamiltonian
formulation (see for  instance \cite{luca,rov}).

The torsion-free derivative operator compatible with
${}^{4}\sigma^{a~A'}_{~A}$ will be denoted
$\nabla$, while the derivative operator associated to the selfdual
connection ${}^{4}\!A_{aA}^{~~B}$ is denoted by ${}^{4}{\cal D}$.
Let $t$ be a function with nowhere vanishing gradient, whose level
surfaces $\Sigma_t$ are diffeomorphic to $\Sigma$.  We denote the
(three dimensional) pull back to $\Sigma_t$ of four dimensional
quantities with the same notation as the four dimensional quantity,
taking away the superscript ''${}^4$'', if present. Let also
$t^{a}$ be a vector
field with affine parameter $t$, \ie~ $t^{a}\nabla_{a}t=1$. Given a
soldering form
${}^{4}\sigma^{a}_{AA'}$ such that each $\Sigma_t$ is spacelike,
take the
normal to $\Sigma_t$ as $n^{a}:~n^{a}n_{a}=-1$. The induced metric
on
$\Sigma_t$ is denoted $q_{ab}:=g_{ab}+n_{a}n_{b}$. Thereby every
tensor field
can be decomposed into its parts on and orthogonal to $\Sigma_t$.
In particular, $t^{a}$
gives us the Lapse and Shift: $t^{a}=Nn^{a}+N^{a}$.
Unprimed $SL(2,\complex)$ spinors, on $M$, and $SU(2)$ spinors, on
$\Sigma$,
can be related as follows. One defines the hermitian
metric in the spinor space, $G^{AA'}:= i\sqrt{2} \, n^{AA'}~
\left(=\bar{G}^{A'A}\right)$,
where $n^{AA'}:= \sigma^{~AA'}_{a} n^{a}$.
The primed $SL(2,\complex)$
spinors define a hermitian conjugation operation on the unprimed
ones through
$\left( \psi^{\dagger} \right)_{A} := G_{A}^{~A'} \bar{\psi}_{A'}$. The
hermitian
conjugate of $\psi_{A}$. The soldering form for $SU(2)$ spinors
\be
\sigma^{a}_{AB} := -i \sqrt{2} \, {}^{4}\sigma{}_{~(A}^{a~~A'}
n_{B)A'},
\ee
is hermitian
$\left(\sigma^{a}_{AB}\right)^{\dagger}= \sigma^{a}_{AB}$, and
traceless
$\sigma^{a~A}_{~A}=0$; it provides, locally, an isomorphism
between
$T\Sigma$ and
second rank, trace-free, hermitian spinors $\gamma_{AB}$. In
particular
$q^{ab}=-{\rm Tr} \left(\sigma^a \sigma^b \right)$.

The configuration variables can be taken as
$A_{aA}^{~~B}$
and $\psi^{A}$,
whereas the correspondent canonically conjugated momenta are $-
i\sqrt{2}
\, \widetilde{\sigma}^{a~B}_{~A} := -i\sqrt{2} \, (\sigma)\,
\sigma^{a~B}_{~A}$
and $ \widetilde{\pi}_{A}:= -i(\sigma)
\left(\psi^{\dagger}\right)_{A}$.
The action of the Einstein-Weyl system becomes:
\bea
S_{\rm Einstein-Weyl} &=& \int dt\int_{\Sigma_{t}} \dthree
 \widetilde{\pi}_{A}\  {\cal L}_{t} \psi^{A}
+ {\rm Tr}\left[ \left(-i\sqrt{2}\, \widetilde{\sigma}^{b}\right)
                   {\cal L}_{t} A_{b} \right] +
\nonumber \\
                      & &
+ \left( t\cdot {}^{4}\!A^{AB}\right) \widetilde{G}_{AB}
+ N^{a} \widetilde{V}_{a}
+ \ut{N} \widetilde{\widetilde{S}}  .
\eea
Where ${\cal L}_t$ is the Lie derivative along $t^{a}$. Here
\bea
\widetilde{G}_{AB} & = & - i\sqrt{2}\, {\cal
D}_{b}\widetilde{\sigma}^{b}_{~AB}
                         + \widetilde{\pi}_{(A} \psi_{B)} \nonumber\\
\widetilde{V}_{a} & = & - i\sqrt{2}\, {\rm
Tr}\left(\widetilde{\sigma}^{b}
                        F_{ab}\right)
                        - \widetilde{\pi}_{A}{\cal D}_{a}\psi^{A}
\nonumber\\
\widetilde{\widetilde{S}} & = & -{\rm
Tr}\left(\widetilde{\sigma}^{a}
                              \widetilde{\sigma}^{b} F_{ab} \right)
+ i \sqrt{2}\, \widetilde{\sigma}_{~A}^{a~B}\, \widetilde{\pi}_{B}
{\cal D}_{a}\psi^{A}.
\label{eq:constraints}
\eea
These are the Gauss, vector and scalar constraints, respectively, for
the
Einstein-Weyl system as given in terms of Ashtekar variables.
Finally, the theory is defined by the conventional reality conditions
which pick the real sector of the phase space to which
classical
general relativity belongs. We will not discuss reality conditions in
this paper.

The symplectic structure one arrives at is
\bea
\left\{ \widetilde{\sigma}^{a}_{~AB}(x), A_{b}^{~CD}(y) \right\}
& = & -\frac{i}{\sqrt{2}}\delta^3 (x,y) \delta_{b}^{~a}
      \delta_{(A}^{~~C} \delta_{B)}^{~~D}, \nonumber \\
\left\{ \widetilde{\pi}_{A}(x), \psi^{B}(y) \right\}
& = & - \delta^3 (x,y) \delta_{A}^{~B}.
\label{eq:symplectic}
\eea
The relation between these canonical variables and the vectorial
Ashtekar variables (see \cite{review}), namely the densitized triad
field $\tilde E^{ai}(x)$ and the $SO(3)$ connection $A_a^i(x)$, is
given by
\be
\widetilde{\sigma}^{a}_A{}^B(x) = {-i \over\sqrt{2}}\ \tilde
E^{ai}(x)\ \sigma_{iA}{}^B, \ \ \ \
A_{aA}{}^{B}(x)  = {-i\over 2}\
A_a^i(x) \  \sigma_{iA}{}^B
\ee
where $i=1,2,3$ and $\sigma_i$ are the Pauli matrices.

With this simplectic structure, the constraints can be
interpreted as generators of gauge transformations, spatial
diffeomorphisms and the very dynamics. This is shown, together
with the fact that they form
a Poisson algebra, in [Ashtekar \etal] by smearing the constraints as
\bea
G_{T} &\equiv& - \int_{\Sigma} \dthree\, T^{BA}\widetilde{G}_{AB}
               \label{eq:smgaussconstraint} \\
D_{v} &\equiv& \int_{\Sigma} \dthree\, \left[ v^{a}\widetilde{V}_{a}
-
               v^{a} A_{a}^{~BA} \widetilde{G}_{AB} \right]
               \label{eq:smdiffconstraint} \\
H_{\ut{N}} &\equiv& i\sqrt{2}\, \int_{\Sigma} \dthree\, \ut{N}
                    \widetilde{\widetilde{S}}\label{eq:smhamconstraint}.
\eea
$T^{AB},v^{a},\ut{N}$ are arbitrary test fields.

\subsection{Classical fermion paths}

Non-perturbative quantum General Relativity is constructed in terms
of the loop variables \cite{carlolee}
\bea
T[\alpha] & := & {1/over 2}\ U_{A}^{~A}[\alpha] \\
{T}^{a}[\beta](s) & :=  &\sqrt{2}\  U_{A}^{~B}[\beta] (s) \
                     \widetilde{\sigma}^{a~A}_{~B}(\beta(s)).
\eea
We indicate loops by greek letters.  A loop is here a closed
continuous
piecewise analytical curve in $\Sigma$, $\alpha: S_1
\rightarrow
\Sigma$;  and $s\in [0,1]$ is the parameter along the loop: $\alpha: s
|\!\rightarrow
\alpha^a(s)$ (We identify the values $s$ and $s+1$).     We indicate
by $U_{A}^{~B}[\alpha](s)$ the parallel
transport $SL(2,\complex)$ matrix of the Ashtekar connection
around the loop $\alpha$, starting from the parameter value $s$;
that is, the
path order exponential of the line integral of the connection around
the loop.
\be
	 U_{A}^{~B}[\beta] (s) := P exp\{\int_s^{s+1} ds\,
{d\alpha^a(s)\over ds}\, A_{aA}{}^B\} .
\ee

In order to construct the extension of the loop representation to
fermions, we want to generalize these loop variables to the
presence of the spinor field. We want to define objects invariant
under the action of the Gauss constraint.   Loop-like variables
in theories with connections involving  fermions, have been dealt
with by Gambini and collaborators \cite{gambini} for  Yang-Mills,
and by Kim
\etal~ \cite{kim}, for 2+1 gravity. For earlier related ideas, see
\cite{wilson,loops}.  We follow here the lines of \cite{rslattice}.
Let us consider
piecewise differentiable continuous {\it open\ } curves in $\Sigma$.
As we said in the introduction, we will call these open lines as
loops,  certainly with an abuse of terminology;  and we denote
them too by means of greek letters:  $\alpha: (0,1) \rightarrow
\Sigma;$ and
$\alpha: s |\!\rightarrow \alpha^a(s)$.   We denote as $\alpha_i$ and
$\alpha_f$
the initial and final point of the loop, namely:
\bea
\alpha_i   & := & \alpha(0),\nonumber  \\
\alpha_f   & := & \alpha(1).
\eea
If $\alpha_f=\beta_i$, we denote the open line obtained joining
$\alpha$ and $\beta$ as  $\alpha\cdot\beta$; that is
\f
[\alpha\cdot\beta](s)    :=  \alpha(2s),\ \ { \rm if }  \  s \in [0,1/2]
\ff \f
[\alpha\cdot\beta](s)   :=  \beta(2s-1),\ \ {\rm if}\  s\in [1/2,1].
\ff

We then define
\bea
X[\alpha] & := & \psi^{A}(\alpha_i)\  U_{A}^{~B}[\alpha]\
\psi_{B}(\alpha_f) \\
{Y}[\alpha] & := & \widetilde{\pi}^{A}(\alpha_i)\  U_{A}^{~B}[\alpha]\
\psi_{B}(\alpha_f).
\label{eq:observables}
\eea
$X$ and ${Y}$, are parametrized by open curves. They are defined
as path integral exponentials of the Ashtekar connection along these
curves, with spinors variables attached to the end points.   They are
clearly $SU(2)-$invariant.  Other important properties of the $X$ and
${Y}$ variables are the following.
\begin{enumerate}
\item They are invariant under a positive derivative monotonic
reparametrization of the open loops.
\item $X$ is invariant under inversion of the open loop, $X[\alpha^{-
1}] = X[\alpha]$. This important property follows from the fact the
fermions are Grassman variables. In fact:
\bea
X[\alpha^{-
1}]&=&\psi^A(\alpha^{-1}_i)
U_A{}^B[\alpha^{-1}]\psi_B(\alpha^{-1}_f)\nonumber \\
&=&\psi^A(\alpha_f)
U_A{}^B[\alpha^{-1}]\psi_B(\alpha_i)\nonumber \\
 & = & -\  \psi^A(\alpha_f)U_{AB}[\alpha^{-
1}]\psi^B(\alpha_i)\nonumber \\
 & = & +\  \psi^A(\alpha_f)U_{BA}[\alpha]\psi^B(\alpha_i)\nonumber
\\
 & = & -\ \psi^B(\alpha_i)U_{BA}[\alpha]\psi^A(\alpha_f)\nonumber
\\
 & = & +\ \psi^B(\alpha_i)U_B{}^A[\alpha]\psi_A(\alpha_f)\nonumber
\\
 & = & +\  \psi^A(\alpha_i)U_A{}^B[\alpha]\psi_B(\alpha_f)\nonumber
\\
 & = & \  X[\alpha]
\eea
In the third and sixth line we have used the spinor index property
$\xi^A\rho_A=-\xi_A\rho^A$. In the fourth line we have used the
parallel propagator property $U_{AB}[\alpha] = -U_{BA}[\alpha^{-1}]$
(recall that if the parallel propagator $U_A{}^B[\alpha]$ is the
identity, then $U_{AB}[\alpha] = \epsilon_{AB}$). In the fifth line,
we have switched the two fermions, gaining a minus sign due to
their Grassmanian character.
\item Retracing identity. As their closed loops counterparts, the
fermionic loop variables $X$ and $Y$ satisfies the retracing and
spinor identities that follows from their being defined in terms of
parallel propagators of an $SU(2)$ connection. For instance, if
$\beta_f=\gamma_i=\delta_i$,  then
$X[\beta\cdot\gamma\cdot\gamma^{-1}\cdot\delta]
= X[\beta\cdot\delta]$.
\item Spinor identity. The following notation is useful. Let $\alpha$
be an (oriented) open line. We define
\f
            \alpha^A \  := \  \psi^B(\alpha_i)\  U_B{}^A[\alpha].
\ff
Then, if $\alpha_f=\beta_f$, we can write
\f
   X[\alpha\cdot\beta^{-1}] = \alpha^A\beta_A
\ff
Now, consider a point $p$ where 4 lines terminate, that is
$\alpha_f=\beta_f=\gamma_f=\delta_f=p$.  There are three possible
ways of connecting these four lines to form two gauge invariant $X$
variables:
\bea
X[\alpha\cdot\gamma^{-1}]  X[\delta\cdot\beta^{-1}] & = &
\alpha_A\gamma^A \delta_B\beta^B = \alpha^A\beta^B
\gamma^C \delta^D \epsilon_{AC}\epsilon_{DB}\\
X[\alpha\cdot\delta^{-1}]  X[\gamma\cdot\beta^{-1}] & = &
\alpha_A\delta^A \gamma_B\beta^B = \alpha^A\beta^B
\gamma^C \delta^D \epsilon_{AD}\epsilon_{BC}\\
X[\alpha\cdot\beta^{-1}]  X[\gamma\cdot\delta^{-1}] & = &
\alpha_A\beta^A \gamma_B\delta^B = \alpha^A\beta^B
\gamma^C \delta^D \epsilon_{AB}\epsilon_{CD}.
\eea
By using the fundamental spinor identity, which is at the root of the
Mandelstam relations
\f
\epsilon_{AB}\epsilon_{CD}+\epsilon_{AD}\epsilon_{BC}+\epsilon_{
AC}\epsilon_{DB}=0,
\ff
we have the fermion version of the spinor identity, namely
\f
 X[\alpha\cdot\gamma^{-1}]  X[\delta\cdot\beta^{-1}]
+X[\alpha\cdot\delta^{-1}]  X[\gamma\cdot\beta^{-1}]
+X[\alpha\cdot\beta^{-1}]  X[\gamma\cdot\delta^{-1}]
\ =\ 0\ .
\ff
Similarly, it is simple to derive the identitity that refers to the
intersection between the open loop $\alpha\cdot\beta$ and the
closed loop $\gamma$, where $\alpha_f=\beta_i= \gamma_i=
\gamma_f$:
\f
X[\alpha\cdot\beta]\ T[\gamma]=X[\alpha\cdot\gamma\cdot\beta]+
X[\alpha\cdot\gamma^{-1}\cdot\beta]
\ff
\item Fermionic (Grassmann) identities. Since Grassman variables
anticommute, and since the fermion field has only two components,
if we multiply three or more fields in the same point we obtain zero.
It follows that we can have products of $X$ variables with at most
two coinciding hands. Thus, for instance, if
$\alpha_i=\beta_i=\gamma_i$, then $X[\alpha]X[\beta]X[\gamma]=0$.
\item Gauge observables algebra. Finally, the most important
property of the $X,Y,T,T^a$ variables is that their Poisson algebra
closes.  A direct computation yelds
\bea
\left\{X[\beta],X[\alpha]\right\}
& = & 0
\nonumber
\\
\left\{{Y}[\beta],X[\alpha]\right\}
& = & \delta^{3}(\alpha_{f},\beta_{i})\, X[\alpha\cdot \beta]
                + \delta^{3}(\beta_{i},\alpha_{i})\, X[\alpha^{-1}\cdot
\beta]\nonumber
\\
\left\{{Y}[\alpha],{Y}[\beta]\right\}
& = & \delta^{3}(\alpha_{f},\beta_{i})\, {Y}[\alpha\cdot \beta]
                - \delta^{3}(\alpha_{i},\beta_{f})\,  {Y}[\beta\cdot \alpha]
\label{eq:xyalgebra}
\eea
Whereas the nonvanishing brackets with the $T$ variables are
\bea
\left\{{T}^{a}[\gamma](s),X[\alpha]\right\}
&=& i \Delta^{a}[\gamma(s),\alpha] \sum_{\mu=\pm 1} \mu\,
      X[\alpha\#_s \gamma^{\mu}] \nonumber \\
\left\{{T}^{a}[\gamma](s), {Y}[\alpha]\right\}
&=& i \Delta^{a}[\gamma(s),\alpha] \sum_{\mu=\pm 1} \mu\,
      \widetilde{Y}[\alpha\#_s \gamma^{\mu}].
\label{eq:txyalgebra}
\eea
\end{enumerate}

As in the pure gravity, one may define also higher order observables,
which will have simple quantum operators associated.  For instance
we can insert "hands" into the $X$ variable, and so on. In order to
treat the dynamics, the following variable will be particularly
useful.
\f
Y^a[\alpha](s) \ :=\ \pi^A(\alpha_i) \,
U_A{}^B[\alpha](0,s)\, \tilde\sigma^a{}_B{}^C(\alpha(s))\,
U_C{}^D[\alpha](s,1)\, \psi_D(\alpha_f),
\label{Ya}
\ff
where the notation $U_A{}^B[\alpha](s,t)$ indicates the matrix of the
parallel trasport along $\alpha$ from $s$ to $t$.
This variable is quadratic in the momenta and will play a role
analogous to the role of the $T^2$ variable in pure gravity, which is
also quadratic in the momenta. Indeed, we will use it for defining
the hamiltonian constraint.

Finally, we may introduce a further small simplification in the
notation: if the context is clear, we may write $X[\alpha]$ as
$T[\alpha]$. Namely we can use the same notation, $T[\alpha]$, to
indicate the the trace of the holonomy of $\alpha$ if $\alpha$ is a
closed loop, and to indicate the parallel propagator along $\alpha$
sandwhiched between two fermion fields if $\alpha$ is open.

\section{QGD: kinematics}

Let us now begin the construction of the quantum theory.  Following
the philosophy described in refs. \cite{review,carlolee,isham}, we
look
for a quantum representation of the classical loop algebra.
Rather than simply writing the rappresentation, it is perhaps more
instructive to use the Loop Transform, introduced in ref.
\cite{carlolee}, as
an heuristic device to help us in this task. Thus we first
consider a "Schr\"odinger-like", rapresentation of the quantum
Einstein-Weyl theory \cite{sft}. We consider
functionals $\Psi[A,\psi]$ on the configuration space, and we define
the canonical coordinates operators $A$ and $\psi$ as multiplicative
operators, and the corresponding momentum operators as functional
derivative operators
\bea
\hat{\tilde\sigma}^a{}_A{}^B(x) &=&  {1\over\sqrt{2}}\
{\delta\over\delta
A_a{}^A_B(x)},  \\
\hat{\tilde\pi}_A(x) &=& \imath {\delta\over\delta \psi^A(x)}.
\eea
Since we are using this construction only as an heuristic tool to find
a possible form for the fermion loops operators, we are not
particularly concerned with mathematical rigor here.  We focus on
loop states in this representation. These are defined as follows.
Given a closed loop $\alpha$, we write
\f
\Psi_\alpha[A,\psi]= U[\alpha]_A{}^A.
\ff
Given an open loop $\alpha$, we write
\f
\Psi_\alpha[A,\psi]= \psi^A(\alpha_i)\, U[\alpha]_A{}^B\,
\psi_B(\alpha_f).
\ff
Given an arbitrary collection $\beta$ of a finite number of open or
closed loops $\alpha_1,\alpha_2, ... , \alpha_n$, we write
\f
\Psi_\beta = \Psi_{\alpha_1} \Psi_{\alpha_2} ... \Psi_{\alpha_n}.
\ff
Note that we follow here the usual convention of denoting loops
(open or closed) and multiple loops (namely collections of a finite
number of loops) by means of the same notation, that is greek
letters from the begining of the alphabet.   The idea of the loop
representation is to take the $\Psi_\beta$ states as an
overcomplete basis of quantum states.  We thus introduce the Loop
Transform \cite{carlolee} as follows.
\f
\Psi[\beta] := \int {\cal D}A {\cal D}\psi\ \Psi_{\beta}[A,\psi]\
                    \Psi[A,\psi]. \label{eq:ltransform}
\ff
For a rigorous mathematical definition of these kind of integrals see
the recent work of Ashtekar and Isham, and Ashtekar and
Lewandowsky \cite{ai}
Here $\Psi[A,\psi]$ represents a generic wave functional in the
connection representation, and $\Psi[\beta]$ represents its Loop
Transform.  The novel features entering here
come from the inclusion of fermions; states containing no fermions
are described exactly as in the pure gravity case \cite{carlolee}.
The transform (\ref{eq:ltransform}) has a well defined meaning on
the
lattice \cite{b}, where the $\Psi_{\beta}$ states are the Wilson-Susskind
states. In the lattice case it is possible to show that the transform
defines a unitary transformation to a new basis in the Hilbert space
of the theory.

Eq.(\ref{eq:ltransform}) suggests that we may look for a
representation of the full Weyl-Einstein loop algebra on a space of
functionals of multiple loops, where the multiple loops are sets of
closed as well as open loops.  Then the transform gives us
immediately the action of the $X$ and $Y$  operators in the loop
representation
\bea
\left(\widehat{X}[\alpha] \Psi\right) [\beta]
&=& \int {\cal D}A {\cal D}\psi\,
     \Psi_\beta[A,\psi] \left(\widehat{X}[\alpha] \Psi\right) [A,\psi]
\nonumber\\
&=& \int {\cal D}A {\cal D}\psi\, \Psi_\beta[A,\psi]
\Psi_\alpha[A,\psi] \Psi [A,\psi] \nonumber\\
&=& \int {\cal D}A {\cal D}\psi\, \Psi_{\beta\cup\alpha}[A,\psi]
\Psi [A,\psi] \nonumber\\
&=& \Psi [\beta\cup\alpha]\,;
\eea
where the union operaton $\cup$ of set theory is well defined
between the multiple loop $\beta$ and the single open loop $\alpha$.
Thus we have
\f
\widehat{X}[\alpha] \Psi[\beta] = \Psi[\beta\cup\alpha].
\label{eq:xloop}
\ff
Note that this is the essentially the same action as the action of
$T[\alpha]$ in pure gravity. Using the notation suggested as the end
of the previous section, we can write
\f
\widehat{T}[\alpha] \Psi[\beta] = \Psi[\beta\cup\alpha]
\ff
for open as well as for closed $\alpha$'s.

As far as ${Y}$ is concerned, we have
\bea
\left(\widehat{{Y}}[\alpha]\Psi\right) [\beta]
&=& \int {\cal D}A {\cal D}\psi\, \Psi_\beta[A,\psi]
\left(\widehat{{Y}}[\alpha] \Psi\right) [A,\psi] \nonumber\\
&=& \int {\cal D}A {\cal D}\psi\, \left(\widehat{{Y}}[\alpha]
\Psi_\beta\right) [A,\psi]\ \Psi[A,\psi] \nonumber\\
&=& \int {\cal D}A {\cal D}\psi\, \left(\psi_{B}(\alpha_f)
U_{A}^{~B}[\alpha]
\frac{\delta~~~~}{\delta\psi^{A}(\alpha_i)} \Psi_\beta
\right) [A,\psi]   \
    \Psi[A,\psi] \nonumber\\
&=& \sum_{\beta_f} \delta^{3}(\alpha_i,\beta_f)\,
\Psi[\alpha\cdot\beta]
+\sum_{\beta_i} \delta^{3}(\alpha_i,\beta_i)\,
    \Psi[\beta\cdot\alpha^{-1}].
\eea
Here $\sum_{\beta_i}$ indicates the sum over all the initial points
of
the open loops in the multiple loop $\beta$, and $\sum_{\beta_f}$
indicates the sum over all the final points of the open loops in the
multiple loop $\beta$. We introduce the following notation. We write
$\beta_e$ to indicate any end point of the multiple loop $\beta$. If
$\beta_e=\alpha_i$, the notation $\beta\cdot_e\alpha$ indicates
the multiple loops obtained by attaching $\alpha_i$ with $\beta_e$.
Thus we have
\f
\widehat{{Y}}[\alpha]\,\Psi[\beta]=
\sum_{\beta_e} \delta^{3}(\alpha_i,\beta_e)\
\Psi[\beta\cdot_e\alpha].
\label{eq:yloop}
\ff
In words, the action of the operator $Y[\alpha]$ is simply to attach
the open loop $\alpha$ to any open end that happens to be in the
point
$\alpha_i$.

Next, we supplement (\ref{eq:xloop}), (\ref{eq:yloop}) with the usual
quantum $T$-variables
in the loop representation \cite{carlolee,review}. The computation
of the
quantum commutation relations of the entire set is then
straigthforward.  The result is that the set of operators $\hat X,
\hat Y, \hat T, \hat T^a$ provides a representation of the  classical
algebra (\ref{eq:xyalgebra}), (\ref{eq:txyalgebra})
and the classical $\cal T$--algebra.

We can also naturally introduce quantum operators corresponding to
higher order loop variables \cite{carlolee}. As their pure gravity
counterparts, these
have a quantum algebra that reduces to the corresponding classical
Poisson algebra in the limit in which the Planck constant goes to
zero. We write here the quantum operator corresponding to the $Y^a$
variable defined above, since it will be used in the construction of
the
Hamiltonian
\bea
Y^a[\alpha](s) \ \Psi[\beta] & = & \sum_{\beta_e}
\delta^3(\beta_e,\alpha_i)
\int_\beta \! dt  \dot\beta^a(t) \delta^3(\alpha(s),\beta(t))
\nonumber \\
 & &
\left[\Psi(\alpha**^+_{e,t}\beta)+\Psi(\alpha**^-_{e,t}\beta)
\right] =
\nonumber \\
 & = & \sum_{\beta_e}
\delta^3(\beta_e,\alpha_i)
\int_\beta \! dt  \dot\beta^a(t) \delta^3(\alpha(s),\beta(t))
\nonumber \\
 & &
\sum_{q=\pm}\Psi(\alpha**^q_{e,t}\beta)
\label{quantumYa}
\eea
Here we have indicated by $\alpha**^+_{e,t}\beta$ and $\alpha**^-
_{e,t}\beta)$ the two loops obtained by joining the $\beta_e$ end
point of $\beta$ with the point $\alpha_i$, and rerooting the
intersection $\alpha(t)=\beta(s)$ in the two possible ways.  Note the
plus sign between the two terms, which will play an important role
in what follows. The expression (\ref{quantumYa}) can be obtained
for instance from the loop transform. To determine the correct
overall coefficient and sign, an accurate computation with the SU(2)
index algebra is needed. Note however, that the relative plus sign
between the two terms in parenthesis is forced by symmetry, since
neither of the two can be preferred.

\subsection{Diffeomorphisms and diff-invariant states}

The classical vector constraint generates spatial diffeomorphisms
when
acting on gauge invariant objects.  This is also true in the
quantum theory providing the correct ordering of the vector
constraint
quantum operator is chosen.  Precisely as in the pure gravity theory,
there
are several equivalent ways for reaching this conclusion:
\begin{enumerate}
\item As suggested by Isham \cite{isham}, the vector constraint can
simply be
defined
as the generator of the natural action of the diffeomorphism group
on the
space of the open and closed loops. The commutator algebra of these
generators
among
themselves and with  all the other operators in the theory, then,
reproduces
the corresponding classical Poisson algebra. This is a sufficient
condition
to insure that the classical limit of the quantum theory that we are
constructing reproduces the classical theory we started from. Since
the correct classical limit is
the {\it sole\ } requirement we have on the theory, the quantum
diffeomorphism constraint defined in this way
represents a consistent quantization of its classical counterpart.
\item We can use the transform, and define the loop representation
vector constraint opererator as the transform of the vector
constraint operator in the representation that diagonalizes $A$ and
$\psi$.
\item We can express the classical vector constraint in terms of
loop
variables, as a suitable limit of a sequence of these variables. The
corresponding quantum constraint is then defined as the limit of the
corresponding quantum loop operators.
\end{enumerate}
As in pure gravity, it is not difficult to show that these different
strategies
yeld the same quantum diffeomorphism constraint, and that this is
can be
expressed as follows. For every diffeomorphism $\phi\in{\rm
Diff}[\Sigma]$
\f
		\Psi[\alpha]\ = \ \Psi[\phi\cdot\alpha]
\ff
where $[\phi\cdot\alpha](s):=\phi(\alpha(s))$. The general solution
of the
diffeomorphism constraint is the given by the loop functionals
constant
along the orbits of the action of  the Diff group  on the loop space,
namely, they
are given by
\f
		\Psi[\alpha]\ = \ \Psi[K(\alpha)],
\ff
where $K(\alpha)$ is a generalized knot class, that is, an equivalent
class
under diffeomorphisms of sets of graphs formed by open and closed
lines.

For every generalized knot class $K$, we can define a corresponding
quantum
state $\Psi_K$ as the characteristic function of the class. We will
also use a Dirac notation $\Psi[\alpha]=\langle\alpha|\Psi\rangle$,
and denote the state $\Psi_K$  as $|K>$. Thus
\bea
  \langle\alpha|K\rangle & = & 1\ \ \  {\it if}\ \  K=K(\alpha) ,
\nonumber \\
  \langle\alpha|K\rangle & = & 0\ \ \  {\it otherwise} .
\eea
Let us begin here some preliminary investigation
of the structure of the ensemble of quantum states $|K>$.

Consider a fixed class $K$. Let $\alpha$ be one of the
(diff-equivalent)
multiple loops that belongs to $K$. Let $\alpha$ be composed by $c$
closed
loops and $o$ open loops. There are $2o$ end points $\alpha_e$ in
$\alpha$.
We distinguish the end points as simple or doubles.  An end-point
$\alpha_e$
is simple is there is no other end point $\alpha_e'$ in $\alpha$ such
that
$\alpha_e=\alpha_e'$.  It is double otherwise. Note that there may
not be
"triple" end points, because of the Pauli principle.  Let $S$ and $D$
be the
number of simple and double end points. And $N=S+2D=2o$ be the
number of end points.

Consider the points $i$ in the immage of $\alpha$ such that one or
more
than one of the following is true:
\begin{enumerate}
\item $\alpha$ is non injective in $i$,
\item $i$ is an endpoint,
\item $\alpha$ is non differentiable in $i$;
\end{enumerate}
we denote these points $i$ as {\it generalised intersections}, or,
simply as
{\it intersections}.  We assume that the number of these
intersections is
finite, and we denote this number as $I$.  Given an intersection $i$,
we
assume there is only a finite number of components of $\alpha$
coming out of it. We denote
this number as $m_i$, and we call it the {\it order\ } of the
intersection.
Intersections of order 1 are single end points. Intersections of order
2 are
either double end points or kinks along a loop. Intersections or order
3 are
single end points that fall over a loop. Intersections of order 4 are
either
crossings of two loops, or a double endpoint that falls over a loop,
and so on.  We call the intersections of order one {\it free\ }
end-points.

Consider an intersection $i$ of order $m_i$ in a loop $\alpha$.
Let $\alpha_j(s)$, with $j=1...m_i$, and $\alpha_j(0)=i$,  be the
$m_i$ lines (components of the loop $\alpha$) that come out from
$i$.
Let $\vec l_j,\ \ j=1...m_i$ be the $m_i$ tangents of $\alpha_j(s)$
in
$i$.
Let then $\vec l^{(k)}_j$, where $k$ is a positive integer, the $k-$th
derivative of the $j-$th component of the loop in $i$.    For instance,
if
$m_i=1$, and $i=\alpha(0)$, then $\vec l_1= d\\vec alpha(s)/ds$,
and
$\vec l^{(2)}_1= d^2\vec \alpha(s)/ds^2$, and so on.

The vectors $\vec l^{(k)}_j$ transform among
themselves under reparametrization of $\alpha$ and under
diffeomorphisms.  Consider the space of all possible intersections
of order $m_i$.  Consider to of these intersections as equivalent if
they can be transformed into each other by diffeomorphisms or
reparametrizations. Denote the space of the resulting equivalence
classes as the moduli space of the
intersection of order $m_i$. The moduli space of an intersection of
order $n<5$ is discrete; not so, in general, for larger $n$.  However,
the moduli space is always finite dimensional.  Let $a^{(m_i)}_i$ be
a collection of parameters that coordinatizes the moduli space of
the $i$ intersection, as well as characterizing the rootings of
$\alpha$ through the $i$ intersection.

Finally, let ${\cal K}^c_P$ be a discrete index that labels the braids
with $P$ (ordered) open hands and $c$ closed loops. We can then
(over-) characterise a
quantum state as (see ref. \cite{RSphysham}):
\f
| N, I, D; \ \    a^{m_1}_1 ....  a^{m_I}_I ; \ \  {\cal
K}^o_{{\scriptscriptstyle \sum}_i m_i} \rangle
\label{notation}
\ff
where, we recall, $N$ is the number of end-points, $I$ is the number
of intersections, $D$ is the number of double end points, ${m_1} ....
 {m_I}$
are the orders of the $I$ intersections, $a^{(m_1)}_1 ....  a^{(m_I)}_I$
are
the moduli space parameters of the intersections and ${\cal
K}^o_{\sum_i m_i}$ is the discrete topological class of the braid
obtained by deleting all intesection points from the loop. At the end
of the paper we will describe some simple generalized knot classes
and make use of this notation.

 \section{The hamiltonian constraint has a simple action}

Our last and main task is to deal with the dynamics, which is
contained in the
hamiltonian constraint.  (The Hamiltonian constraint of the pure
gravity Loop Representation was discussed in
\cite{carlolee,gambini,hamc}. For a comprehensive and critical
discussion of the various approaches see ref. \cite{brug}.)
We shall perform this task in two stages.
In the present section we introduce a simple and naive non-
regularized definition of the quantum hamiltonian constraint.   This
is not really satisfactory because it does not allow us to control the
divergences of the theory.  However, we think it is usefull to present
this non-regularized version of the dynamics first, because it
allows one to appreciate the striking simplicity of the geometrical
action of the Weyl-Einstein hamiltonain constraint, which otherwise
could improperly appear as an improbable product of the
technicalities of the regularization procedure.  In the next two
sections, we will transform the formal esxpressions we obtain here
in a more solid result.

Thus, we begin by defining the hamiltonian constraint by using the
simplest
procedure: we define it in the connection representation and
we transform it to the Loop Representation by using the Loop
Transform \cite{carlolee}.  We choose the ordering in which the
momenta are
always to the right of the configuration variables. Taking into
account the Grassmann character of the fermionic variables, the
Hamiltonian constraint, smeared against a test (inverse density)
scalar $ \ut{N}(x) $ is
\bea
\widehat H[\ut{N}] & = & - \int_{\Sigma}\dthree\ \ut{N}\  {\rm Tr}\,
\left(F_{ab} \widehat{\widetilde{\sigma}}^{a}
                              \widehat{\widetilde{\sigma}}^{b} \right)
- i\, \sqrt{2}\   \widehat{\widetilde{\sigma}}_{~A}^{a~B}\,
{\cal D}_{a}\psi^{A}\ \widehat{\widetilde{\pi}}_{B}
\nonumber \\
& = & - {1\over 2}\ \int_{\Sigma}\dthree\ \ut{N}(x) \
F_{ab}^A{}_B(x) {\delta\over\delta A_{aB}^C(x)}
                              {\delta\over\delta A_{bC}^A(x)}
\nonumber \\
& & - i \ {\cal D}_{a}\psi^{A}(x)\
 {\delta\over\delta A_{bB}^A(x)}\
{\delta\over\delta \psi^{B}(x)}
\label{ham}
\eea
In order to compute the Loop transform of this operator, we have to
compute its action on the kernel of the Loop Transform, that is, on
the basis loop states $\Psi_\alpha[A,\psi]$.  We need to compute
\f
	\widehat H[\ut{N}] \ \Psi_\alpha[A,\psi].
\ff
If $\alpha$ is formed by closed loops alone, then
$\Psi_\alpha[A,\psi]$ is independent from $\psi$ and therefore the
second term in (\ref{ham}) does not act.  It follows immediately
that the action of $\widehat H[\ut{N}] $ on the closed loop states is
fully equivalent to the action of the pure gravity hamiltonian
constraint.  Let us then assume $\alpha$ is a single open loop.  A
straigth forward calculation gives
\bea
\lefteqn{\widehat{H}[\ut{N}] \ \Psi_\alpha[A,\psi] = } \nonumber \\
& & - \int_{0}^{1} dt\,\, \int_{0}^{1} ds\, \ut{N}(\alpha(t))
    \delta^3(\alpha(s),\alpha(t)) \dot{\alpha}^{a}(s)\,\,
\dot{\alpha}^{b}(t)\ \nonumber \\
& &
    \psi^{G}(\alpha_i)\, U_{G[F}(0,t) F_{ab}^{~~FB}(\alpha(t))
 U_{B]H}(t,1)\,
    \psi^{H}(\alpha_f) \nonumber \\
& & -\, \int_{0}^{1}
    ds \, \ut{N}(\alpha_i) \delta^3(\alpha(s),\alpha_i)\,
    \dot{\alpha}^a(0) \, {\cal D}_a \psi^A (\alpha_i)
    U_{AE}[\alpha]\, \psi^E (\alpha_f) \nonumber \\
& & + \, \int_{0}^{1}
    ds \, \ut{N}(\alpha_f) \delta^3(\alpha(s),\alpha_f)\,
    \dot{\alpha}^a(1)\, \psi^{D} (\alpha_i)
    U_{DA}[\alpha]\,  {\cal D}_a \psi^A(\alpha_f).
\label{eq:regham}
\eea
We immediately see the difficulty in this equation: the three-
dimensional delta functions are integrated only against two line
integrals, leaving a divergent factor.  As we said, let us disregard
this infinity for the moment.

We now recall from ref.\cite{carlolee} that, if we disregard
divergences, the
action of the Hamiltonian operator on a pure loop state can be
written as
\f
\widehat{H}[\ut{N}] \ \Psi_\alpha = {\cal S}[\ut N] \Psi_\alpha
\label{hamshift}
\ff
where ${\cal S}$, denoted as the "Shift operator" is a simple
operator {\it acting on the loop argument\ }, as
\f
{\cal S}[\ut{N}] \Psi_\alpha :=  \int_{0}^{1} ds \   \int_{0}^{1} dt \
\ut{N}(\alpha(s))  \dot\alpha^a(s) \delta^3(\alpha(s),\alpha(t))
{\delta\over\delta \alpha^a(t)} \ \Psi_\alpha.
\label{shiftop}
\ff
If $\alpha$ has no self-intersections, and up to a divergent factor
$k$,  the Shift operator becomes simply
\f
{\cal S}[\ut{N}] \Psi_\alpha = k \lim_{t->0} \Psi_{\alpha_{t\ut{N}}}
\label{shift}
\ff
where
\f
\alpha^a_{t\ut{N}}(s)
:= \alpha^a(s) + t\ut{N}(\alpha(s))  \dot\alpha^a(s)
\ff
The action of the Shift operator has thus a very simple geometrical
interpretation: it shifts the loop ahead along its tangent.  Clearly, it
sends a smooth closed loop into itself, while it deforms a loop with
kinks or intersections.  This simple action is one of the ways in
which one can (formally) understand the well known result that loop
states corresponding to smooth non intersecting loops are in the
kernel of the Hamiltonian constraint.   The fact that the action of
the hamiltonian constraint on the kernel of the Loop Transform can
be expressed in terms of an operator acting on the loop variable
enables us to interpret this operator as the operator that represents
the Hamiltonian constraint in the Loop Representation
\cite{carlolee}.
Thus in pure gravity the Hamiltonian constraint can be simply
expressed as the Shift operator.

How does the picture change if we include the fermions ?  The
striking result that we have mentioned in the introduction is that
the picture {\it does not change at all\ } with the introduction of
fermions: Equations (\ref{hamshift}),
which expresses the Hamiltonian as the Shift operator still holds.
The shift operator is still given by equation (\ref{shiftop}),
where now we also allow the loops to be open, or in the absence of
intersections, by equation  (\ref{shift}).  Indeed, by computing the
action of ${\cal S}[\ut{N}]$, as defined in equation  (\ref{shiftop}) on
an {\it open\ } loop state, we get precisely the right hand side of
equation
(\ref{eq:regham}).   What happens is that the fermion term in the
classical hamiltonian constraint give rise to the second and third
term in equation (\ref{eq:regham}) and these terms are precisely the
terms that "move" the two fermions at the end of the loop in the
correct direction !

Thus we have the following result:

\begin{itemize}

\item In pure gravity the action of the hamiltonian constraint on
Loop Space can be expressed as the action of the Shift operator
(\ref{shiftop}), which simply shifts non-intersecting loops along
their own tangent.

\item By applying this {\it same\ } geometrical operator on {\it
open\ } loop states, we have the action of the Weyl-Einstein
hamiltonian constraint.

\end{itemize}

This result was first obtained by one of us in ref.\cite{hugo}.  The
present paper represents an effort to clean up this result from the
difficulties due to the divergent factor $k$.

We consider the fact that the coupled Einstein-Weyl equations can
be coded in the extremely simple Shift operator as a striking fact.
We have been deeply puzled by this result, and we do not see any
simple way of interpreting it.  We have not been able to find any
reason for which this result could be understood in terms of the
classical fermion dynamics.   Its simplicity seems to us an
indication of something, but we have not been able to decode the
indication.  We will return to this discussion in the conclusion
section.   For the moment, let us study how we can free the result
from the divergences.

\section{QCD: dynamics}

\subsection{Coupling a clock field}

Following the ideas introduced in the introduction, we now modify
our point of view, and consider a richer theory:  We couple a scalar
field to the
Einstein-Weyl theory.    Our aim to use the scalar field in order to
define a physical internal time, or clock-time, in terms of which
we can study physical evolution.

We have two independent motivations for chosing this roard.    First
this procedure allows us to unravel the physics of the general
covariant quantum theory, otherwise hidden in the frozen-time
formalism, as discussed in detail for instance in
refs.\cite{observable,qobservable}.     Second,
this is a way to overcome the divergence difficulties we had in the
previous section. Indeed, we have learned from the experience in
pure Quantum GR that non-diffeomorphism-invariant operators tend
to be ill-defined in a generally covariant quantum field theory, while
all the diffeomorphism invariant operators that we have been able to
construct have good finiteness properties \cite{weave}.   Thus, we
wish to
replace the Hamiltonian constraint with some diffeomorphism
invariant operator [the Hamiltonian constraint, being a scalar
(density) is diffeomorphism covariant, not diffeomorphism
invariant].   As shown in ref.\cite{RSphysham}, the coupling of a
scalar clock field
and the replacement of the hamiltonian constraint with an
hamiltonian is a way to acheive this result.   The hamiltonian that
generates the evolution in the clock time is a diffeomorphism
invariant quantity and replaces the hamiltonian constraint.

We refer to \cite{RSphysham} for the details of the scalar field
construction and the
gauge fixing that allows to define the hamiltonian.  We simply recall
here the main idea, so that this paper could be idependently read.
Physical quantum states are represented (say in the connection
representation) as functionals $\Psi[A,\psi]$  of the spacial fields
$A(\vec x), \psi(\vec x)$, satisfying the Wheler-DeWitt equation.  As
well known, the time coordinate $t$ disappears from this frozen
time formalism.  In theory, the disappearence of the coordinate
time is not a problem, since the observables of the theory must be
4-dimensional general covariant {\it anyway\ }, and thus, in
particular, must be independent from $t$ anyway.  Examples of these
4-dimensional general covariant observables are given by the
invariant distance $d_p$ of the solar system planets from the Earth,
seen as a function dependent from, say, the invariant distance $d$ of
the Earth from the sun, or as a function of the proper time (on the
Earth) lapsed from some initial fixed event. These observables are
manifestly coodinate invariant, and thus independent from $t$.   In
practice, however, it is notoriously too
difficult to write the dynamical equations of GR {\it directly\ } in
terms of coordinate invariant quantities: in classical GR we almost
always work with coordinate dependent quantities, and extract
coordinate invariant predictions only after the dynamics has been
fully worked out in a particular gauge.  For instance, we study the
motion of the planets, including perhaps emitted gravitational
radiation, in some arbitrary coordinate system (with some arbitrary
coordinate time $t$: $d_p(t)$ and $d(t)$), and only when the
dynamics has been solved we compute coordinate invariant
quantities $d_p(d)$, which can be compared with astronomical
observations.  The quantum frozen time formalism, however, does
not allow us to work with quantities dependent on the fictitious
arbitrary coordinate time $t$, and thus makes the dynamical
analysis particularly cumbersome:  what one should do is to view the
physical states $\Psi[A,\psi]$ as coding the quantum evolution of
any variable in terms of any other variable.  In general this is not
easy.

The problem can be simplified by studying a version
of the theory in which there is a simple quantity to be taken as the
independent variable; that is, in which we know from scrath which
variable we want to use as the "clock variable".  We thus introduce a
scalar field ${T}(\vec x,t)$, and we decide to study the evolution of
the gravitational and fermions degrees of freedom, as they evolve
in the value of ${T}$. If $A(\vec x, t), \psi(\vec x, t), {T}(\vec x, t)$
is a solution of the field equations, we extract information invariant
under a time coordinate transformation by solving $t$ with respect
to ${T}$, and substituting the resulting $t(\vec x,{T})$ in $A$ and
$\psi$:  We get the two functions
\bea
A(\vec x, { T}) &=& A(\vec x, t(\vec x,{T})),
\nonumber \\
\psi(\vec x, {T}) &=& \psi(\vec x, t(\vec x, {T})),
\label{**}
\eea
which are invariant under coordinate time reparametrization.
Equivaletly, we choose a (physical) coordinate system in which
${T}(\vec x, t ) = t$.   Of course, this cannot be done in general. It can
only be done for certain solutions of the field equations, and for
certain regions of spacetime.  Thus, we are now beginning to use a
formalism that holds only within a certain regime, and breaks down
in approaching the limits of this regime - for instance, when the
derivative ${\partial T \over \partial t}$ vanishes.

In the quantum theory, the frozen time formalism is defined in
terms of the
functionals $\Psi[A,\psi,T]$  of the spacial fields $A(\vec x),
\psi(\vec x), T(\vec x)$.  We can interpret these states as giving the
amplitude for a $A(\vec x), \psi(\vec x)$ configuration {\it at the
given configuration $T(\vec x)$ of the clock field\ }, and to interpret
the Wheeler-DeWitt equation as an evolution equation in the
multifingered time $T(\vec x)$. We can then further fix the
multifingered $T(\vec x)$ time as a constant function $T(\vec x) =
T$, (we keep the same letter $T$ to indicate also the real number
$T$, beside the function $T(\vec x)$),  and restrict $\Psi[A,\psi,T]$
({\it functional\ } of the field $T(\vec x)$) to $\Psi[A,\psi](T)=
\Psi[A,\psi,T]|_{T(\vec x)=T}$ ({\it function\ } of the real numeber
$T$), without any loss of information. The state $\Psi[A,\psi](T)$
expresses the quantum amplitude for the evolution in $T$ of the
fields $A(\vec x, { T}), \psi(\vec x, {T})$ defined above. Moreover, if
$\Psi[A,\psi,T]$ satsfies the Wheeler-DeWitt constraint, it is shown
in \cite{RSphysham} that $\Psi[A,\psi](T)$ satisfies the equation
\f
i\hbar{\partial\over\partial T}\ \Psi[A,\psi](T) =
\widehat{H}\ \Psi[A,\psi](T),
\label{shroedy}
\ff
where the operator $\widehat{H}$ will be defined in a moment.
We can view equation (\ref{shroedy}) as a genuine Schr\"odinger
equation which evolves in the time $T$.

More precisely, we can express the Einstein-Weyl-scalar-field
theory in a canonical (gauge-fixed) form, in terms of the
configuration
variables $A_{aA}^{~~B}$ and $\psi^{A}$, the gauge and
diffeomorphisms constraints
\bea
\widetilde{G}_{AB} & = & - i\sqrt{2}\, {\cal
D}_{b}\widetilde{\sigma}^{b}_{~AB}
                         + \widetilde{\pi}_{(A} \psi_{B)} \nonumber\\
\widetilde{V}_{a} & = & - i\sqrt{2}\, {\rm
Tr}\left(\widetilde{\sigma}^{b}
                        F_{ab}\right)
                        - \widetilde{\pi}_{A}{\cal D}_{a}\psi^{A}
\eea
and the {\it hamiltonian\ } (as opposed to hamiltonian constraint)
\f
H  =  \int d^3x\ \sqrt{
-{\rm Tr}\left(\widetilde{\sigma}^{a}\widetilde{\sigma}^{b} F_{ab}
\right)
+ i \sqrt{2}\, \widetilde{\sigma}_{~A}^{a~B}\, \widetilde{\pi}_{B}
{\cal D}_{a}\psi^{A}}.
\label{ham}
\ff
It is shown in \cite{RSphysham} that the solutions of this theory,
$A(\vec x, { T}),
\psi(\vec x, {T})$, are related to the solutions of the full
Einstein-Weyl-scalar-field theory via equation (\ref{**}).
We can define the
kinematics of the quantum theory and treat the diffeomorphism
constraint exactly as in the previous sections, but now we do not
have a quantum hamiltonian constraint, but rather a Sch\"odinger
equation (\ref{shroedy}) and a quantum hamiltonian $\hat H$, which
is the quantum
operator corresponding to the classical variable $H$ given in
(\ref{ham}).

The rest of this section is devoted to the construction of the
quantum operator $\widehat H$. This time we will not be
content with formal manipulations of divergent expressions, and we
require a somehow higher level of rigor.

\subsection{Definition of the regularized hamiltonain}

In this section we derive the key techincal result of the paper,
namely the explicit regularized action of the Hamiltonian on the knot
classes; the section is quite long and technical - the final result is
summarized in the following section 5.3.
Following ref.\cite{RSphysham}, our first step is to definine a
regularised classical
hamiltonian in terms of loop variables.  We introduce a fictitious
background flat metric and a preferred set of coordinates in which
this metric is euclidean; in the following everything is written in
those coordinates and we use the euclidean metric to define norms
of vectors and raise and lower vector indices. It will be our task,
later, to show that the operator we define is independent from the
regularization background metric introduced here.   We write
\f
{H} = \lim_{\stackrel{\scriptstyle L\rightarrow 0
\;\delta\rightarrow 0}{
               A\rightarrow 0\; {\tau}\rightarrow 0}}
H_{L \delta A \tau },
\label{reg}
\ff
\be
H_{L \delta A \tau } = \sum_{I} L^3 \sqrt{- {C}^{A,L,\delta}_{{\rm
Einstein}\; I}
                   - {C}^{L,{\tau},\delta}_{{\rm Weyl}\; I}} \ .
\ee
Here we have partitioned three-dimensional space into cubes of
sides $L$, labelled by the index $I$. The quantity
${C}^{A,L,\delta}_{{\rm Einstein}\; I}$ is the regularized form of the
pure gravity hamiltonian constraint (integrated over the $I-$th
cube), and has been defined in \cite{RSphysham}.

With the purpose of defining the fermion term ${C}^{L,{\tau},
\delta}_{{\rm Weyl}\; I}$, we begin by defining the open loop
$\gamma^\tau_{\vec x,\vec y}$, where $\tau$ is a regularization
parameter (to be taken small), $\vec x$ is a point in space and $\vec
y$ is a vector in the tangent space around $\vec x$. Since we have a
background flat metric we can write expressions as $\vec x + s\vec
y$, for any
real $s$, with obvious meaning.  The open loop $\gamma^\tau_{\vec
x,\vec y}$ is defined as the  (uniformly parametrized) straight line
(in the background metric) that starts at $\vec x$ in the $\vec y$
direction and is long $\tau$.  That is
\f
(\gamma^\tau_{\vec x,\vec y})^a(0)=x^a,\ \ \
(\dot\gamma^\tau_{\vec x,\vec y})^a(s)={\tau\over |\vec y|} y^a.
\ff
Note that
\f
(\gamma^\tau_{\vec x,\vec y})^a(1)=x^a+{\tau\over |\vec y|} y^a
\ff
and
\f
(\gamma^\tau_{\vec x,\vec y})^a({|\vec y|/ \tau})=x^a + y^a.
\ff
By making use of this loop, we define the regularised fermion
hamiltonian constraint by
\bea
{C}^{L,{\tau},\delta}_{{\rm Weyl}\; I}
&=& \frac{1}{L^3} \int_I d^3x\; {C}^{{\tau},\delta}_{\rm Weyl}(\vec
x)\; ,
\\
{C}^{{\tau}, \delta}_{\rm Weyl} (\vec x)
&=&  c_{\tau\delta}
\int\!\! d^3y\; \theta(\delta-|y|) \ {y^a\over |\vec y|}  \ {Y}^{{a}}
\left[\gamma^{\tau}_{\vec x \vec y }\right](|y|/\tau)
  \;.
\label{c}
\\
 c_{\tau\delta} & = & \frac{3}{\tau}
 \frac{1}{{4\over 3}\pi\delta^3}\; .
\eea
Here $\theta(x)$ is the conventional step function, that is the
characteristic function of $R^+$.   Note the three different roles of
the three regularization parameters $L$, $\tau$ and $\delta$: the
parameter $L$ fixes the size of the boxes. As we will see, the
introduction of these boxes will allow us to deal with the square
root.  For every space point $x$, the fermion term of the Hamiltonian
constraint $C_{\rm Weyl} (\vec x)$ is approximated by means of the
loop variable $Y^a$ corresponding to the "small" loop
$\gamma^{\tau}_{\vec x \vec y }$, that starts at $x$.  The parameter
$\tau$ gives the length of the "small" loop.  The direction of this
loop is integrated over ($d^3y$ angular integration). $Y^a$ has a
special point where the hand is inserted.  This point is chosen to be
$\gamma^{\tau}_{\vec x \vec y }(|y|/\tau) = x^a+y^a$.  Thus, also the
position of this point (the hand) is integrated over  ($d^3y$ radial
integration). This smearing of the position of the hand is what
determines the point split of the functional derivatives operators in
the quantum theory. Note that the $d^3y$ integral is restricted (by
the $\theta$ function) to a ball of size $\delta$ around $x$. Thus,
the parameter $\delta$ gives the point splitting separation between
the initial point of the loop and its hand. Note that in order this
definition to make sense we must have  ${\tau}>\delta$.

A straigthforward expansion in $\tau$ and $\delta$ shows that
equation (\ref{reg}) is satisfied, namely that the quantity that we
have defined represents a genuine regularization of the Hamiltonian.
For completness, let us sketch this expansion.  We begin by writing
the regularized expression explicitely, by using the
expression for $Y^a$, given in eq.(\ref{Ya})
\bea
{C}^{{\tau}, \delta}_{\rm Weyl} (\vec x)
& = &  c_{\tau\delta}
\int\!\! d^3y\; \theta(\delta-|y|) \ {y^a\over |\vec y|}  \
\ \pi^A(\gamma^{\tau}_{\vec x \vec y }{}_i)
U_A{}^B\left[\gamma^{\tau}_{\vec x \vec y }\right](0,|y|/\tau)
\nonumber \\
& &
\tilde\sigma^a{}_B{}^C(\gamma^{\tau}_{\vec x \vec y }(|y|/\tau))
U_C{}^D\left[\gamma^{\tau}_{\vec x \vec y }\right](|y|/\tau,1)
\psi_D(\gamma^{\tau}_{\vec x \vec y }{}_f).
\eea
By using the explicit definition of the loop $\gamma^{\tau}_{\vec x
\vec y }$, this becomes
\bea
{C}^{{\tau}, \delta}_{\rm Weyl} (\vec x)
& = &  c_{\tau\delta}
\int\!\! d^3y\; \theta(\delta-|y|) \ {y^a\over |\vec y|}  \
\ \pi^A(\vec x)
U_A{}^B\left[\gamma^{\tau}_{\vec x \vec y }\right](0,|y|/\tau)
\nonumber \\ & &
\tilde\sigma^a{}_B{}^C(\vec x+\vec y)
U_C{}^D\left[\gamma^{\tau}_{\vec x \vec y }\right](|y|/\tau,1)
\psi_D(\vec x+\tau\vec y/|\vec y|).
\eea
We may now expand everything in powers of $\delta$ and $\tau$
around the point $\vec x$. Before doing so, however, let us note what
follows.
Since we have a ($d^3y$) integration over a volume of order
$\delta^3$, and we divide by $\delta^3$ (in $c_{\tau\delta}$) before
taking the limit $\delta\rightarrow 0$, only the term in the
integrand  of zero
order in $\delta$ may survive the limit. We also divide by
$\tau$ and take the limit $\tau\rightarrow 0$, thus only terms of
first order
in $\tau$ survive in the limit.  This means that to the relevant order
we may replace quantities in $\vec x+\vec y$ by quantities in $\vec
x$ (recall the $d^3y$ integration is over a sphere of radius
$\delta$), and quantities in $\vec x+\tau\vec y/|\vec y|$, which is a
distance $\tau$ from $\vec x$ by the first two terms in their Taylor
expansion around $\vec x$.  By doing so, the first of the two parallel
propagators is just replaced by the identity, while the second can be
replaced with the entire parallel propagator along the small loop.
We therefore have, up to terms of order $\delta$ or $\tau$,
\bea
{C}^{{\tau}, \delta}_{\rm Weyl} (\vec x)
&=&  c_{\tau\delta}
\int\!\! d^3y\; \theta(\delta-|y|) \ {y^a\over |\vec y|}  \
\ \pi^A(\vec x)
\delta_A{}^B
\nonumber \\ & &
\tilde\sigma^a{}_B{}^C(\vec x)
\left(\delta_C{}^D + \tau  y^c A_{cC}^D(\vec x) \right)
\left(\psi_D(\vec x)+\tau y^c\partial_c\psi_D(\vec x)\right)=
\nonumber \\
&=&  c_{\tau\delta}
\int\!\! d^3y\; \theta(\delta-|y|) \ {y^a\over |\vec y|}  \
\ \pi^A(\vec x)
\nonumber \\ & &
 \tilde\sigma^a{}_A{}^C(\vec x)
\left[\psi_C(\vec x) +
\tau  y^c \left(
A_{cC}^D(\vec x) \psi_D(\vec x) +\partial_c\psi_C(\vec x)
\right) \right]
\eea
The first term in the square brackets vanishes because $\int d^3y
y^a=0$, and the second term is the covariant derivative. Thus
\f
{C}^{{\tau}, \delta}_{\rm Weyl} (\vec x)
=  c_{\tau\delta}\tau
\left[\int\!\! d^3y\; \theta(\delta-|y|) \ {y^a\vec y^c\over |\vec
y|}\right]  \
\ \pi^A(\vec x)  \tilde\sigma^a{}_A{}^C(\vec x)
D_c\psi_C(\vec x).
\ff
The integration is now immediate
\f
\int\!\! d^3y\; \theta(\delta-|y|) \ {y^a y^c\over |\vec y|} =
3 \ \delta^{ac}\ {4\over 3}\pi\delta^3
\ff
Restoring the explicit form of $c_{\tau\delta}$ we thus obtain
\f
{C}^{{\tau}, \delta}_{\rm Weyl}\ =\   \pi^A \tilde\sigma^a{}_A{}^C
D_c\psi_C +O(\delta)+O(\tau),
\ff
namely the fermion term of the hamiltonian constraint, as we
wanted.

We have shown that the quantity (\ref{c}) provides indeed a
regularized form of the fermion hamiltonian constraint.  We now
come to the quantum theory and define the corresponding regularized
quantum operator
 \bea
\widehat{H}_{L \delta A \tau }
& = & \sum_{I} L^3 \sqrt{- \hat{C}^{A,L,\delta}_{{\rm Einstein}\; I}
                   - \hat{C}^{L,{\tau},\delta}_{{\rm Weyl}\; I}} \ .
\\
\hat {C}^{L,{\tau},\delta}_{{\rm Weyl}\; I}
&=& \frac{1}{L^3} \int_I d^3y\; \hat {C}^{{\tau},\delta}_{\rm Weyl}
    (y)\; , \label{eq:wcube}
\\
\hat{C}^{{\tau}, \delta}_{\rm Weyl} (y)
&=& c_{\tau\delta}\
\int\!\! d^3x\; \theta(\delta-|y|) \ {y^a\over |y|}  \ \hat{Y}^{{a}}
\left[\gamma^{\tau}_{\vec x \vec y }\right](|y|/\tau)
  \;. \label{eq:wreg}
\eea
The operator $\hat{Y}^{{a}}$ is defined in equation (\ref{quantumYa}).
Note that in this regularized operator the two hands of the operator
\cite{carlolee} do not overlap: they are point split.

Let us study the action of the operator that we have defined: in spite
of it apparent complexity, this action will turn out to be relatively
simple. Plugging the explicit definition of $\widehat{Y}^a$
(\ref{quantumYa}) in (\ref{eq:wreg})  gives
\bea
\hat{C}^{L,{\tau},\delta}_{{\rm Weyl}}(\vec x)
\;\Psi[\alpha] & = & {c_{\tau\delta}}
\int\!\!d^3y\;
 \theta(\delta-|y|) \ {y^a\over |y|} \ \sum_{\alpha_e} \
\delta^3(\vec x,\alpha_e)
\nonumber \\
& &
\int_\alpha ds\
\dot\alpha^a(s) \
\delta^3(\gamma^\tau_{\vec x\vec y}(|\vec y|/\tau),\alpha(s))
\ \ \sum_{q=\pm}\Psi\left[\alpha**^q{}_{e,s}\gamma^{\tau}_{xy}
\right]\  .
\eea
We now use the explicit form of the small loop, and we keep only
terms small in $\tau$ and $\delta$.  The first $\delta^3$ function in
the last equation forces the loop $\alpha$ to have an end-point
$\alpha_e$ in $\vec x$. The integration over the small ball  (size
$\delta$) around this point, and the second  $\delta^3$ function, pick
up a second point $\alpha(s)=\vec x+\vec y$ in $\alpha$, close to
$\alpha_e$.\ \  If $\alpha_e$ is a free end point, we can write this
second point (up to the relevant order) as $\vec x+|y|\vec l_e$,
where $\vec l_e$ was defined in the previous section as the tangent
of $\alpha$ in the end-point $\alpha_e$.  In general, however,
$\alpha_e$ needs not be a free end-point; in the general case there
will be several components of $\alpha$ originating from
$\alpha_e$.  Thus $\vec l_e$ takes a finite number of values, and the
radial $d^3y$ integration together with radial part of the second
delta function pick up all these values; note that they turn
out to be proportional to $\vec y$.  The radial $d^3y$ integration is
then straightforward, and we obtain
\f
\hat{C}^{L,{\tau},\delta}_{{\rm Weyl}}(\vec x)
\;\Psi[\alpha] = {c_{\tau\delta}\over L^3} \delta
\  \sum_{\alpha_e} \ \delta^3(\alpha_e,\vec x) \
\sum_{\vec l_e} \sum_{q=\pm}
\Psi\left[[\alpha**^q{}_{e,\delta}
\gamma^{\tau}_{\vec x(\vec x+\delta\vec l)}   \right]\  .
\ff
We introduce the notation
\f
\hat{C}^{L,{\tau},\delta}_{{\rm Weyl}}(x)
\;\Psi[\alpha] ={c_{\tau\delta}\over L^3} \delta
\  \sum_{\alpha_e} \ \delta^3(\alpha_e,x) \
\widehat{\cal F}^{\tau\delta}_e\ \Psi[\alpha]
\ff
\f
\widehat{\cal F}^{\tau\delta}_e\ \Psi[\alpha] =
\sum_{\vec l_e} \sum_{q=\pm}
\Psi\left[[\alpha**^q{}_{e,\delta}
\gamma^{\tau}_{\vec x(\vec x+\delta\vec l)}   \right]\  .
\label{defF}
\ff

Finally, we may come to the Hamiltonian. Let us assume for a
moment that the loop we are dealing with has no intersection nor
kinks, so that we can set the Einstein term to zero.  Inserting our
last result into the Hamiltonian we have
\f
\hat{H}\Psi[\alpha] = \lim_{\stackrel{\scriptstyle L\rightarrow 0
\;\delta\rightarrow 0}{{\tau}\rightarrow 0}}
\sum_{I} L^3 \sqrt{ {c_{\tau\delta}\over L^3} \delta
\  \int_{{\rm cube}\ I} \! d^3x\sum_{\alpha_e} \ \delta^3(\alpha_e,x)
\
\widehat{\cal F}^{\tau\delta}_e\
} \Psi[\alpha] \ .
\ff
For $L$ small enough, and assuming that $\delta<<L$, so that
"boundary effects" of the box can be neglected, every cube $I$
contains only one
end-point, and since $\hat{C}_{\rm Weyl}$ gives zero unless there is
an end-point in $I$, we have
\f
\hat{H}\Psi[\alpha]= \lim_{\stackrel{\scriptstyle L\rightarrow 0
\;\delta\rightarrow 0}{{\tau}\rightarrow 0}}
\sqrt{ \frac{3}{\tau}
 \frac{1}{{4\over 3}\pi\delta^3}{ L^3} \delta }
\sum_{\alpha_e}  \left(\widehat{\cal F}^{\tau\delta}_e
\right){}^{-{1\over 2}} \ \ \Psi[\alpha].
\ff
where we have restored the explicit expression for $c_{\tau\delta}$.

It is now time to study the limits explicitely. Let us first focuss on
the crucial prefactor
\f
C(L,\tau,\delta) = \sqrt{ \frac{3}{\tau}
 \frac{1}{{4\over 3}\pi\delta^3}{ L^3} \delta }=
\frac{3}{2\sqrt{\pi}} \sqrt{\frac{L^3}{\tau\delta^2}}.
\ff
The question we have to address is the finitness of the above limit.
Clearly the result depends on the order in which the limits are taken.
This is precisely what we should have expected: different orders in
which the limit is taken correspond to inequivalent definitions of
the quantum operator. Since all these definitions correspond to the
same classical limit, the choice amount to a choice of different
orderings of the quantum hamiltonian. The question is whether there
is {\it one\ } choice that gives us a finite quantum operator.

Of course we may not confine ourselves to the choice between taking
one first  or another one first of the three limits: we choose any
combination. More
precisely, we may consider the three dimensional $L,\tau,\delta$
space, and study the limit of $C(L,\tau,\delta)$ as we
approach the point $L=0,\tau=0,\delta=0$: this point can be
approached in a variety of alternative ways, not just along one of
the coordinate axis. Let us introduce a parameter $\epsilon$, and
consider a curve $L(\epsilon), \tau(\epsilon), \delta(\epsilon)$ in
this tree dimensional space, such that $L(0)=0, \tau(0)=0,
\delta(0)=0 $. Our problem is to understand wether we can choose
this curve in such a way that the limit
\f
\lim_{\epsilon\rightarrow 0}
C(L(\epsilon), \tau(\epsilon), \delta(\epsilon))
\ff
is finite.  We are not free to choose the curve $L(\epsilon),
\tau(\epsilon), \delta(\epsilon)$ in a completly arbitrary way,
because there is a certain number of conditions that we have
imposed on the regularization parameters along the way. First, of
course, we must have $L>0, \tau>0, \delta>0$. Then, have required
$\tau> \delta$, and, in order to avoid "boundary effects" in the box,
$L > \delta$.  Can all the conditions be satisfied and a finite limit be
obtained ?

The first crucial point to be noted is that powers of lengths cancell
exactly and the quantity $C(L,\tau,\delta)$ is dimensionless.  This is
a necessary condition for having a finite limit. (It was precisely the
fact that we got a divergent quantity with the dimensions of a
length, measured in the background metric, that prevented us from
achieving a background independent renormalization in section (5)).
By
itself, however, this fact does not suffice to garantee a well defined
limit.  We now claim that this limit can indeed be chosen
consistently with all the demands, as follows
\bea
      L(\epsilon) &= & k \epsilon^3 a, \nonumber \\
     \tau(\epsilon) &= & \epsilon \, a, \nonumber \\
    \delta(\epsilon) &= & \epsilon^4 a,
\eea
where $a$ is an arbitrary length, and $k$ is a arbitrary
dimensionless positive number. It is easy to see that all the
requested conditions are satisfied. In the limit, we have
\f
\lim_{\epsilon\rightarrow 0}
C(L(\epsilon), \tau(\epsilon), \delta(\epsilon))=
\frac{3}{2\sqrt{\pi}}
\sqrt{\frac{L^3(\epsilon)}{\tau(\epsilon)\delta^2(\epsilon)}}
= \frac{3k^{-3/2}}{2\sqrt{\pi}} := \lambda^2
\ff
$\lambda$ is a free dimensionless constant that emerges from the
regularization procedure. Thus, the prefactor is finite in the limit.

Thus, we write the action of the hamiltonian as
\be
\hat{H}\Psi[\alpha]= \lambda^2\ \
 \sum_{\alpha_e}
\left(\widehat{\cal F}_e
\right){}^{-{1\over 2}}\ \  \Psi[\alpha]
\ee
where we have introduced the "end-point operator"
\be
\widehat{\cal F}_e \Psi[\alpha] =
\lim_{\epsilon\rightarrow 0}
  \widehat{\cal F}^{\tau(\epsilon)\delta(\epsilon)}_e \ \
\Psi[\alpha].
\label{epo}
\ee
We now examine this "end-point operator", its action, its finiteness
and its transformation properties under diffeomorphisms.

Since $\delta$ (the point splitting distance of the two grasps) goes
to zero much faster than $\tau$ (the length of the added loop, we can
now simply take the $\delta\rightarrow 0$ limit first, and the
$\tau\rightarrow 0$ limit second. Let us consider the
$\delta\rightarrow 0$ limit of $\widehat{\cal F}^{\tau\delta}_e
\Psi[\alpha]$ (with finite $\tau$).  It is easy to see that if the end-
point is free, the action of the operator is simply to add a small
straight line of length $\tau=\epsilon a$ to the end point of the loop,
in the direction of the loop tangent (goign out from the intersection).
If the end-point is not free, the
action of the operator produces one term for each component of
$\alpha$ emerging from the end-point.  The terms corresponding to
the component of $\alpha$ that ends in $\alpha_e$ is again just an
addition of a small straight line to the end-point, emerging from the
end point in the same direction as the original loop; while the other
terms imply the addition of the small loop and also a rerouting
through the intersection. The rerouting pattern can be calculated in
straight forward way from equation (\ref{defF}).

Before taking the limit $\tau\rightarrow 0$, let us now assume that
$\Psi[\alpha]$ is a diffeomorphism invariant state. Thus
$\Psi[\alpha]$ depends on the diffeomorphism equivalence class of
$\alpha$ only.   If $\alpha_e$ is a free end-point, we have then
\be
\lim_{\tau\rightarrow 0}
\widehat{\cal F}_e^{\tau 0} \Psi[\alpha] =
\lim_{\tau\rightarrow 0}
2 \Psi[\alpha**\gamma^\tau_{\alpha_e,\dot\alpha|_e}] =
2  \Psi[\alpha],
\label{free}
\ee
because for small enough $\tau$ the added loop will not interject
any other loop, and the addition of a small line at the end of a loop
does not change the diffeomorphism equivalence class of the loop.

If, on the other way, $\alpha_e$ is not a free end-point, then the
loop $\alpha**\gamma^\tau_{\alpha_e,\vec l_e}$ {\it does\ } belong
to a different knot class than $\alpha$.   For instance, assume the
end
point $\alpha_e$ falls over a smooth component $\beta$ of the loop
$\alpha$. One of the terms in (\ref{defF}) adds a small
straight line to $\alpha$. SInce this in the direction in which
$\alpha$ emerges from the intersection, its action is essentially to
cut away an infinitesimal portion of the loop at the end point. This
can be seen by noticing that we can assume without loosing
generality that $\alpha$ is a straight line at the intersection, and
using the retracing identity, to "retrace" $\alpha$ away from the
intersection along the small added loop. In the limit, we go back to
$\alpha$; however, for every finite value of the regularization
parameter, the knot class has changed: we are in the knot class in
which $\alpha$ {\it does not\ } touch $\beta$.  Thus, by continuity in
the limit we still have the knot class in which $\alpha$ {\it does
not\ } touch $\beta$. The key point, now, is that in any case, since
$\Psi[\alpha]$ is
diffeomorphism invariant, for small enough $\tau$ we have that
$\widehat{\cal F}_e^{\tau 0} \Psi[\alpha]$ becomes {\it
independendent from\ } $\tau$.  Therefore the limit is the limit of a
constant function, and therefore is {\it finite}. The action of the
Hamiltonian, therefore, has opened up the intersection between
$\beta$ and the end point of $\alpha$.

It is clear that the resulting action of $\widehat{\cal
F}_e$ is well-defined on the diffeomorphism invariant states.  Thus,
the operator $\widehat{H}$ is finite and diffeomorphism invariant in
the limit.

If we now reinstate $\hat{C}_{\rm Eintsein} \ne 0$, we have
\be
\hat{H}= \sum_{\stackrel{\scriptstyle {\rm intersections}\;i }{
                        {\rm end-points}\;e}}
\sqrt{\hat{M}_i+\lambda\hat{\cal F}_e}\;,
\label{h}
\ee
where $\hat M$ was constructed in \cite{RSphysham}.  $\hat{H}$ is a
finite operator defined on knot states.  Its action
follows immediately from the construction above.

\subsection{Topological Feynman rules}

It is clear from the discussion above that the action of the
Hamiltonian constraint on the generalized knot classes can be
represented in a fully geometrical way in terms of the action of the
operator on single intersections.  The operator "opens up"
intersections, and rearrange the rootings through the intersections.
The details of this action can be computed from equations
(\ref{defF}\ref{epo}\ref{h}), which we report here for easy reference.  The
finite geometrical action on the loops is defined by the operator
(\ref{defF})
\f
\widehat{\cal F}^{\tau\delta}_e\ \Psi[\alpha] =
\sum_{\vec l_e} \sum_{q=\pm}
\Psi\left[[\alpha**^q{}_{e,\delta}
\gamma^{\tau}_{\vec x(\vec x+\delta\vec l)}   \right]\  .
\ff
This operator attaches to every intersection and to every line
coming out from an intersection two small straight loop, one for
each one of the two possible rearranging of the rootings at the
intersection between the added little loop and the line.  Next, we
take the limit (\ref{epo})
\be
\widehat{\cal F}_e \Psi[\alpha] =
\lim_{\epsilon\rightarrow 0}
  \widehat{\cal F}^{\tau(\epsilon)\delta(\epsilon)}_e \ \
\Psi[\alpha].
\ee
And finally, the Hamiltonian is given by (\ref{h})
\be
\hat{H}= \sum_{\stackrel{\scriptstyle {\rm intersections}\;i }{
                        {\rm end-points}\;e}}
\sqrt{\hat{M}_i+\lambda\hat{\cal F}_e}\ .
\label{h}
\ee
The matrix elements of the operators $\hat{M}_i$
(\cite{RSphysham}) and $\hat{F}_e$
can be directly computed between any two given knot states.  The
calculation amounts in a straightforward exercize in geometry and
combinatorics, starting from the two equations above.   The next
problem is to compute the square root of
the resulting (infinite) matrix.  We expect that the square root can
be computed order by order as the complexity of the knots
considered increase. Work is in progress to compute explicitely the
matrix elements, and thus understand if the structure of the
resulting matrix allows a simple argoritm for extracting the square
root.  The resulting geometrical action represents the equivalent of
the Feynman rules for this diffeomorphism invariant, or (infinite
dimensional) topological, theory.  We suggest to denote them as
topological Feynman rules.

\section{QGD: dynamics}

We are now in the position of describing the general structure of
Quantum Gravitational Dynamics, or QGD, the quantum theory of
gravitationally interacting fermions, evolving in the clock time
defined by a scalar field.

A physical quantum state $|K\rangle$ of the theory
is specified by a generalized knot, namely an open braid $K$ of order
$N$ (with $N$ open end-points, $N$ even), with an arbitrary finite
number $I$ of intersections.  A more accurate notation for these
states is given in equation (\ref{notation}).   The quantum dynamics
is given by the
matrix $\hat H$ in braid space, given in equation (\ref{h}), the
matrix elements of which are computed, order by order,
according to the geometrical and algebraic rules given by equations
(\ref{defF}) and (\ref{epo}), and in
ref.\cite{RSphysham}.  We can interpret the matrix elements of
$\hat H$ as first
order transitions amplitudes in a time dependent perturbation
expansion in the clock time $T$.   In principle, the exponentiation of
the $\hat H$ action gives the full evolution.

\subsection{The simplest states}

For instance, we can start from the simplest state formed by a
single nonself intersecting open line. In terms of the notation
(\ref{notation}), this can be denoted as
\f
| 2, 2, 0; 1^0_{2} \rangle
\ff
where we have indicated the simplest value of ${\cal K}^0_{2}$, a
single line, by $1^0_{2}$.  The moduli space of free open-ends is
clearly formed by a single point, and thus we do not need $a_i$
parameters.

There are two
fermions in this quantum state.  We have that  $\hat{M}_i| 2, 2, 0 ;
1^0_{2} \rangle = 0$ because there are no intersections of kinks in $|
2, 2, 0 ; 1^0_{2} \rangle
$.  On the other side, we have from (\ref{free}) that
\f
\hat{F}_e \ | 2, 2, 0 ; 1^0_{2} \rangle
 =  \lambda^2  | 2, 2, 0 ; 1^0_{2} \rangle,
\ff
so that we get
\f
\hat{H} \ | 2, 2, 0 ; 1^0_{2} \rangle
= 2 \lambda  | 2, 2, 0 ; 1^0_{2} \rangle .
\ff
Therefore $ | 2, 2, 0 ; 1^0_{2} \rangle$ is an eigenstate of the theory,
or equivalently, the time dependent Schr\"odinger quantum state
\f
  | 2, 2, 0 ; 1^0_{2},\ \ T  \rangle =
{\rm exp}\left\{i\lambda\sqrt{{c^5\over\hbar
G}}\ T\right\}
 \ \  | 2, 2, 0 ; 1^0_{2} \rangle
\ff
is a solution of the {\it exact\ } quantum interacting theory. (We
have restored physical units, for clarity.)  Perhaps this state
corresponds to an extremely simple "universe"
in which there are only
two fermions gravitating around each other in the simplest of the
quantum geometries.   It is suggestive to think at this state as a
kind of "atomic" "ground state" of a simple 2-fermions universe.

Next we can consider the generalized knot formed by $n$ non
intersecting copies of the above, and denote it as $ | 2n, 2n, 0 ;
1^0_{2n} \rangle
$. It is then straight forward to see that the time evolution
of this state is given by
\f
 | 2n, 2n, 0 ; 1^0_{2n},\ \ T  \rangle = {\rm exp}\left\{i\ n
\lambda\sqrt{{c^5\over\hbar
G}}\ T\right\}
 |  2n, 2n, 0 ; 1^0_{2n}  \rangle
\ff
and that the corresponding energy eigenstates are
\f
	E_n = n E_1 = n\  \lambda\  \sqrt{{c^5\hbar\over G}}.
\ff
As soon as we consider simple intersecting states, the full
complexity of the operators $\hat{M}_i$ and $\hat{F}_e$ becomes
relevant, and we have non-trivial time evolution.

\subsection{Comments}

Before concluding, we list here a certain number of comments and
considerations, as well as pointing out several
important problems that remain open.

\begin{enumerate}

\item {\it Conservation of particle number.}
 The operator $\hat{F}_e$ defined above acts on end-points by
displacing them, and possibly by changing the associated rootings at
intersections, but never creates or destroys end-points. Since the
operator $\hat{M}_i$ too, conserves the number of end-points, it
follows that the hamiltonian that we have defined conserves the
number of particles.  This is at first surprising, given that in
general there is particle creation from space-time dynamics. But a
more close analysis shows that this conservation is to be expected.
Unlikely the Einstein-Dirac theory, indeed, the Einstein-Weyl theory
does conserves particle number. This can be seen classically from
the fact that the quantity
\be
N\ :=\ \int_{\Sigma} \dthree\  \psi^{A}(x)\ \widetilde{\pi}_{A}(x)
\ee
commutes with all the constraints, including the hamiltonain
constraint  \cite{kim}.  One can immediately define the
corresponding operator (say, using the Loop Transform), which turns
out to be
\be
\hat N\| N, I, D; \ \    a^{m_1}_1 ....  a^{m_I}_I ; \ \  {\cal
K}^o_{{\scriptscriptstyle \sum}_i m_i} \rangle \ =\
N\| N, I, D; \ \    a^{m_1}_1 ....  a^{m_I}_I ; \ \  {\cal
K}^o_{{\scriptscriptstyle \sum}_i m_i} \rangle.
\ee
This confirms the interpretation of the number $N$
of end-points as the particle number. Since $[\hat N,\hat H] = 0 $,
the number of end-points $N$ is a conserved quantum number in
the theory.

\item {\it Particle anti-particle distinction.} The Weyl field theory
describe a particle-antiparticle couple (say a neutrino and its anti-
neutrino).  In the Langrangian
formulation the fermions are described by two complex fields.
Since the action contains only first derivatives,
the phase space has the same dimension as the
space of the lagrangian fields, namely four real dimensions per
point. These give two degrees of freedom, which describe, indeed,
the particle and its antiparticle.  Do the end-points of the loops
represent particles or anti-particles ?   The answer is that the
distinction is not gauge invariant, thus the question is not well
posed in the theory. In flat space one can globally
distinguish particles from antiparticles; but when the Weyl system
is coupled with gravitation, something curious happens: the particle
antiparticle distinction becomes local. Consider two field
excitations in
two different space position, and assume the first is a particle;
then, the fact that the second be a particle or an antiparticle
depends on the parallel transport operator between the two. This is
because the particle and the antiparticle are distinguished by
different directions in the internal spin space, and we can only
compare directions in spin space in different points by using the
connection.   Since the particle anti-particle distinction is gauge
dependent, in a gauge-independent formulation there is not way to
distinguish particles from anti-particles. This is why the end-points
of the loop represent at the same time both kinds of excitations.

\item {\it Regime of validity of the formalism, and complex energy
eigenvalues.} This is an important feature of the clock field
formalism that we must be discussed in detail. The
formalism cannot be used in {\it any\ } regime of the system. This is
already obvious at the classical level: Consider an arbitrary solution
of the field equations $A(\vec x,t), \psi(\vec x,t), T(\vec x,t)$: in
general it is {\it not\ } possible to invert $T(\vec x,t) \rightarrow
t(\vec x, T)$ globally.  Thus, we certainly cannot use $T$ as a time
variable for every solutions of the field equations and for every
spacetime region. On the other side, {\it there are\ } solutions and
spacetime reions where we can make the inversion. Consider an
initial configuration of the $A(\vec x),
\psi(\vec x), T(\vec x)$ fields and their time derivatives $\dot
A(\vec x), \dot \psi(\vec x), \dot T(\vec x)$ on a given space like
surface.  This defines a point in phase space.  Assume that $\dot
T(\vec x)<0$ in a spacial region $R$. There will be a time interval
$\Delta t$ for which  $\dot T(\vec x)$ will remain positive in $R$.
More precisely, we can determine a region $\Gamma$ in phase space,
and a corresponding spacetime region $S$, such that for any initial
condition in $\Gamma$,  $\dot T(\vec x)$ is positive in $S$.  We
shall say that the gravitational-fermion-scalar field system is in
the  {\it clock regime\ } in $S$ if the initial conditions are in
$\Gamma$, namely if  $\dot T(\vec x)$ is positive all over $S$.   By
definition, we can perform the inversion $T(\vec x,t) \rightarrow
t(\vec x, T)$ in the region in which the system is in the clock
regime.
Thus, as far as the classical theory is concerned, the formalism
makes sense only in this regime.   The same is true in the quantum
theory.  The quantum formalism that
we have constructed is meanigfull in the clock regime.

A paradigm
for this construction can be found in the quantum system of two
uncoupled simple harmonic oscillator variables $g(t), f(t)$, if we fix
the total energy $E$ and
decide to never consider the evolution in the external clock time $t$,
but rather use one of the two variables, say $f$ as internal clock;
namely if we decide to ask questions concerning what is the position
$g$ of the first oscillator, when the second is in $f$. We obtain  the
classical evolution $g(f)$ by inverting $f(t)\rightarrow t(f)$ and
defining $g(f) = g(t(f))$. We can also do the same in quantum
mechanics (see ref. \cite{model}, where the example is worked out
in detail).
However, along any orbit there is a point in which the internal time
variable $f$ "comes back" (in $t$), and therefore we obtain a non
unitary evolution operator in $f$. The physical interpretation of this
non-unitarity is clear:  there is "no system" anymore for $f$
arbitarry large.

The formalism reflect this fact, both classically and quantum
mechanically, in the form of the hamiltonian.  The hamiltonian that
evolves the two oscillators in the external time $t$ is
$p_g^2+p_f^2+g^2+f^2$.  The hamiltonian that evolves the systemin
the internal time $f$ is easilly obtained solving for $p_f$ (see
ref.\cite{model})
\f
H(f) = \sqrt{E-p_g^2-g^2-f^2}
\label{model}
\ff
(where $E$ is the total energy of the system in the $t$ time). This
(time dependent) hamiltonian generates the evolution equations for
$g(f)$. The important point to note is that the hamiltonian becomes
immaginary for large $f$.  This simply signal the fact that it doesn't
make sense anymore to evolve $g(f)$ in $f$. If we want to continue
the evolution by using an internal time, we have to choose another,
distinct, internal time, and "patch" the evolution.  Note that this
does
not mean that the formalism that evolves in $f$ is inconsistent or
incorrect: it simply means that is has a certain domain of validity.
The same holds in quantum mechanics. Indeed, it was shown in
ref.\cite{model} that
the quantum hamiltonian corresponding to  (\ref{model}) is self-
adjoint when suitably restricted to an ($f-$dependent) region of the
Hilbert space, but develops immaginary eigenvalues if applied
outside this region.

Similarly, we expect that the the hamiltonian that we have defined
will also have immaginary eigenvalues.  These simply signal that
twe are going out from the domain of validity of the formalism,
namely from the {\it clock\ } regime.  We are trying to use evolution
in $T$ to describe the gravitational field in regions where the $T$
fields fails to be monotonic.   Explicitely, this possibility can be
easilly traced back to the classical hamiltonian constraint.  Roughly
speaking, since the form of this constraint is $\Pi^2+C_{\rm
Einstein}+C_{\rm Weyl} = 0$ ($\Pi$ being the scalar field
momentum) and since in order $\dot T$ to change sign $\Pi$ must
vanish, it follows that the save region, is where $C_{\rm
Einstein}+C_{\rm Weyl}>0$, which is of course a sufficient condition
for the Hamiltonian we have definined, $H=\int\sqrt{C_{\rm
Einstein}+C_{\rm Weyl}}$ to be real.  Thus, imaginary eigenvalues of
$\hat H$ signal that we are exiting the regime of validity of the
formalism we have developed here: the object we have chosen as
clock is running backward.   We must therefore exclude from the
state space of QGD, as formulated here, the graphs that are
eigenvalues of $\hat H$ witha an eigenvalue that is not a real
positive number.

In particular, all the vacuum solutions of pure quantum GR that
where previously found lie outside the clock regime. They are
eigenstate of $\hat H$ whith vanishing eigenvalue. Classically, the
vacuum solutions of the theory are only compatible with $\Pi=0$,
namely $\dot T=0$, which clearly indicates that we are outside th
eregime in which we can take the scalar field as a good clock.

We do not consider this necessary restriction of the formalism as a
serious
limitation. Our long term aim is to develop a usable theory that can
be employed, at least in principle, to describe Planck scale
measurements and the Planck scale evolution of quantum geometry.
We would be very content of having a sensible general covariant
field theory that correctly describe this physics in the regime in
which whatever we are using as a clock keeps behaving as a clock.

\item {\it Scalar product.}  One of the weak points of the Loop
Representation is given by the fact that a complete and consistent
definition of the scalar product is not yet available. The
conventional wisdom is that once physical observables on the
physical state space have been constructed, the scalar product is
determined by the requirement these physical observables be
self-adjoint.  The present work is a step in this direction.  The (real
eigenvalues) states of the hamiltonian $\hat H$ must form,
if the formalism is consistent, an orthogonal basis.  Thus, working
out explicitely the eigenstates of $\hat H$ in knot space should at
the same time lead to a partial definition of the scalar product. For
a discussion of the definition of the scalar product for fermion in
functional representations, see \cite{sft}.

\item {\it Taking limits on knot space.}  Finally, let us discuss a
subtle point in the definition of $\hat H$, which we percieve as the
most delicate and potentially problematic point in the construction
above.  We refer to the different way in which the $\delta
\rightarrow 0$ and the $\tau \rightarrow 0$ limit have been dealt
with, when dealing with knot states.

To focus the point, let us consider a model example. Consider the
space $C[R]$ of the continuous functions $f(x)$ on the real line.
Consider the closure $D$ of the space $C[R]$, say in the
pointwise topology, such that $D$ contains also piecewise
continuous
functions as the step function $\theta$ defined by
\f
\theta(x)=1 \  \ {\rm if}\ x>0\ \ ; \theta(x)=0  \ \ {\rm otherwise.}
\ff
Now define the linear functional $k$ on $C[R]$ as follows:
$(k,f):=lim_{x\rightarrow 0} f(x)$, and assume you want to extend
$k$ from $C[R]$ to $D$.  There are two possible strategies: one is to
keep the definition
\f
(k,f)=\lim_{x\rightarrow 0} f(x).
\ff
The other is to note that an equivalent definition of $k$ on $C[R]$ is
$(k,f)=f(0)$, and thus to define
\f
(k,f)=f(0).
\ff
According to the first definition
\f
(k,\theta) = 1,
\ff
according to the second
\f
(k,\theta) = 0.
\ff

We are in a similar situation when we need to study the action of
the hamiltonian $\hat H$ on knot states.  In quantum mechanics,
operators are often defined on dense subspaces of the state space.
For instance we begin by defining the momentum operator in
Schr\"odinger mechanics not on the full $L_2$ state space but on the
dense subspace of the differentiable functions; then we can extend
it.  The Hamiltonian $\hat H$ that we define in this paper contains a
certain number of limiting procedures. We may first rigourously
define it on a suitable restricition of the space of the loop
functionals.
For instance we may assume that the loop functionals are continuous
in all the deformations that we consider. $\hat H$ is well defined on
this space.  Then, however, we want to consider the action of $\hat
H$ on the knot states. These are not continuous in the deformations
that we consider and thus we need to define a suitable extension of
the operator.  At this point we have a choice that essentially
reflects the choice we described in the simple example above.  As
far as we understand, this choice, if not dictated by internal
consistency, is again part of the quantization ambiguities as the
ordering of the dynamical operators.

The important point we want to make here is not that a choice of the
extension has been made in computing the action of $\hat H$ on the
knot states, but that {\it two different\ } choices have been made
for the two limits $\delta \rightarrow 0$ and $\tau \rightarrow 0$.
In fact, as far as the $\delta \rightarrow 0$ is concerned, we have
assumed that we should first take the limit, and then consider the
extension of the action of the operator to diffeomorphism invariant
states; while as far as the $\tau \rightarrow 0$ limit is concerned,
we
have assumed that we should first extend the the operator to
diffeomorphism invariant states, and then take the limit.

This choice is not completely arbitrary: $\delta$ must go to zero
faster that $\tau$, and, if we take away the fake dimensions added
by the integration, we see that the first significative term, which is
the one that we are considering, is of order zero in $\delta$ and of
first order in $\tau$. Ths means that already at the classical level
what we are doing is precisely considering a function
$f(\delta,\tau)$ and picking up terms of the form
${\partial\over\partial \tau} f(\delta,\tau)|_{\delta=\tau=0}$. Thus
it
is not completely unreasonable that this difference gets translated
in the different ways in which the two limits are taken on loop
space: roughly, we are "really" looking at the $\delta=0$ point, and
we are "really" looking at the limit in the first order expansion in
$\tau$.  However, these are
hand-waving justifications of our choice.  Untill a well-defined
calculus on Loop Space is constructed \cite{ai}, we do not see a
way to
transform these tentative explorations into solid mathematics.   Our
only real justification at this point, if any, is the hope that the
(finite) structure we are constructing be internally consistent and,
perhaps, related to Nature.

\end{enumerate}

\subsection{What next}

The next step in the construction of the theory should be to compute
explicitely matrix elements of $\hat{M}_i$ and $\hat{F}_e$ (see
eq.(\ref{h})), and to understand whether there is a direct algoritm
for
extracting the square root.  If this can be done, the theory is
essentially at the stage where the evolution of physical states can
be described.

As noted above, the scalar product is partially fixed by the
construction itself.  The energy and the particle number are
conserved observables. There are other observables in the theory
that one may consider, and evolve, as the area observable discussed
in references \cite{weave,carloobserves}.  A crucial test for the
consistency of the scheme developed here is, as was noted in
\cite{RSphysham}, whether the second order term of the time
dependent perturbation expansion develops divergences.

\vskip1cm

We are deeply indebited with Lee Smolin for ideas, criticism, and
encouragement. This work was partially supported by the NSF Grant
PHY-9311465.


\begin{thebibliography}{99}

\bibitem{carlolee} C. Rovelli and L. Smolin, Phys. Rev.  Lett.
61, 1155 (1988);  Nucl. Phys. B133 (1990) 80.

\bibitem{weave}
A. Ashtekar, C. Rovelli, L. Smolin, Phys Rev Lett 69 (1992) 237

\bibitem{gravitons}
J Iwasaki, C. Rovelli, : ``Gravitons as Embroidery on the
Weave",   International Journal of Modern Physics D1, 533
(1993);  ``Gravitons from loops: non-perturbative
loop-space
quantum gravity contains the graviton-physics approximation",
submitted to {  Classical and Quantum Gravity}.

\bibitem{carloobserves}C. Rovelli, {\it A generally
covariant quantum field theory and a prediction on
quantum measurements of geometry},
{  Nuclear Physics B405}, 797 (1993).

\bibitem{RSphysham}
Rovelli C and Smolin L 1993 {\em "The physical hamiltonian in non-
perturbative quantum gravity}"  Pittburgh University Preprint (1993)
gr-qg ?.

\bibitem{review} C. Rovelli, Classical and Quantum Gravity, 8 (1991)
1613.

\bibitem{abhayreview}
A. Ashtekar, {\it Non-perturbative canonical gravity}.
 Lecture notes prepared in collaboration with R. S. Tate.
(World Scientific, Singapore,1991).

\bibitem{review-ls}
L. Smolin, {\it Recent developments in nonperturbative
quantum gravity} in the Proceedings of
the 1991 GIFT International Seminar on Theoretical Physics:
{\it Quantum Gravity and Cosmology}, (World Scientific,
Singapore, in press).

\bibitem{observable}
 C. Rovelli: ``What is observable in classical and quantum
gravity?'',  {   Classical and Quantum Gravity 8}, 297 (1991).

\bibitem{qobservable}
C. Rovelli: ``Quantum reference systems'', {Classical
and Quantum Gravity 8}, 317 (1991).

\bibitem{tradition}
A. Einstein, {\it \"Uber die spezielle und die allgemeine
Relativit\"atstheorie\ },  (Vieweg, Braunshweig, 1920); P. G.
Bergmann,
Helv. Phys. Acta Suppl. 4, 79 (1956);   J. Stachel, Einstein search for
general covariance 1912-
1915 in Einstein and the History of General Relativity eds. D.Howard
and J. Stachel, Einstein studies, Volume 1, Birkhouser, Boston, 1989.
B. S. DeWitt,  in {\it Gravitation, An Introduction to
Current Research} ed. L. Witten (Wiley, New York,1962).

\bibitem{gambini}R. Gambini and A. Trias, Phys. Rev. D23
(1981) 553,  Lett. al Nuovo Cimento 38 (1983) 497;
Phys. Rev. Lett. 53 (1984) 2359; Nucl. Phys. B238, 436 (1986);
B278 (1986) 436; R. Gambini, L. Leal and  A. Trias, Phys. Rev. D39
(1989) 3127;   R. Gambini, Phys. Lett.
B 255 (1991) 180.  R. Gambini and L. Leal, {\it Loop
space coordinates, linear representations of the diffeomorphism
group and knot invariants} preprint, University of Montevideo, 1991.
{\it SU(2) QCD in the path representation}, University of Montevideo
preprint 1993.

\bibitem{diracsciama} Dirac Sciama

\bibitem{nelson} Nelson

\bibitem{ashtekar}
A. Ashtekar.
 New variables for classical and quantum gravity.
 Physical Review Letters 57, 2244--2247 (1986).
New {H}amiltonian formulation of general relativity.
  Physical Review  D36, 1587--1602 (1987).

\bibitem{abhayfermions}
A. Ashtekar, J.~D. Romano, and R.~S. Tate.
 New variables for gravity: Inclusion of matter.
 Phys.Rev.D  D40, 2572--2587 (1989).

\bibitem{wilson}  K. Wilson, Phys. Rev. {\bf D10}, 247 (1974);
J. Kogut, L. Suskind, Phys. Rev. {\bf D11}, 395 (1975); L. Susskind {\it
Coarse grained quantum chromodynamics} in {it Les Houches XXIX
Weak and  Electromagnetic Interactions at High Energy} ed. R. Balian
and C.H. Llewellyn-Smith  (Amsterdam:North Holland).

\bibitem{hugo}
H. Morales-Tecotl. Ph.D. thesis 1993. Unpublished.


\bibitem{time}
C. Rovelli, in {\it Conceptual Problems of Quantum Gravity} ed.
A. Ashtekar and J. Stachel, (Birkhauser,Boston,1991);
Phys. Rev. D 42 (1991) 2638; 43 (1991) 442.

\bibitem{hypotesis}
C. Rovelli: ``Time in quantum gravity: an hypothesis",
{  Physical Review D43}, 442  (1991).

\bibitem{model}
C. Rovelli : "Quantum mechanics without time: a model" {
Physical
Review D42}, 2638 (1991); ``Quantum evolving constants", {
Physical
Review D44}, 1339 (1991).

\bibitem{clock}
K. Kuchar and C. Torre, Phys. Rev. D43 (1991) 419; D44 (1991)
3116.  L. Smolin, in the Brill Feschrift Proceedings,
ed. B-l Hu, T. Jacobson, (CUP, Cambridge,1993).  K. Kuchar,   in the
Brill Feschrift Proceedings {\it op. cit.}.

\bibitem{ashtekar}A. V. Ashtekar,  Phys. Rev. Lett.
57, 2244 (1986) ; Phys. Rev. D 36 (1987) 1587.

\bibitem{Esposito-Morales-Kim} esposito morales kim

\bibitem{Esposito-Morales} esposito morales

\bibitem{action}
T. Jacobson and L. Smolin.
 The left-handed spin connection as a variable for canonical gravity.
 Physics Letters  B196,  39--42 (1987).
J. Samuel.
 A {L}agrangian basis for {A}shtekar's reformulation of canonical
  gravity.
 { Pram{\=a}na-J Phys.} 28,  L429-L432 (1987).
T. Jacobson and L. Smolin.
 Covariant action for {A}shtekar's form of canonical gravity.
 Classical and Quantum Gravity  5,  583--594 (1988)

\bibitem{tedj} T. Jacobson.
 Fermions in canonical gravity.
 Class. and Quantum Grav.\  5(10):L143--L148, October 1988.

\bibitem{luca}
A.~Ashtekar, L.~Bombelli and O.~Reula, In: {\it Analysis, Geometry
and Mechanics: 200 Years after Lagrange}, eds. M. Francaviglia and D.
Holm (North Holland, Amsterdam, in press).

\bibitem{rov} C. Rovelli, {\it An introduction to canonical Quantum
Gravity. Lecture notes\ }, 1992, Unpublished.

\bibitem{kim} Kim

\bibitem{loops} S. Mandelstam, Ann. Phys. 19 (1962) 1.  A. M.
Polyakov, Phys. Lett. {\bf 82B}, 247 (1979); Nucl. Phys. {\bf B164},
171 (1979).  Yu.M. Makeenko, A.A. Migdal Phys. Lett. {\bf 88B}, 135
(1979).   Y. Nambu, Phys. Lett. {\bf 80B}, 372 (1979); F. Gliozzi, T.
Regge, M.A. Virasoro, Phys. Lett. {\bf 81B}, 178 (1979); M. Virasoro,
Phys. Lett. {\bf 82B}, 436 (1979);  A. Jevicki, B. Sakita, Phys. Rev.
{\bf D22}, 467 (1980); B. Sakita, ``Collective field theory'' in ``Field
theory in elementary particles'' ed. B. Kursunoglu and A. Perlmutter,
Plenum Publishing Coprporation, 1983.

\bibitem{rslattice}
C. Rovelli, L. Smolin: ``Loop representation for Yang Mills theories:
Lattice''.  Roma preprint (1989).

\bibitem{bruglat} B Br\"ugmann, Phys Rev D 43, 566 (1991).



\bibitem{isham} C. J. Isham, in {\it Relativity, Groups and Topology
II}, ed B.S. DeWitt and R. Stora (Amsterdam, Elsevier).


\bibitem{sft} For a review of functional representations of quantum
field theory (including fermions), see J H Yee in "Recent
developments in field theory", ed J E Kim, Mineum Publishing Co
1992.  Functional representations for fermions have been studied by
R. Floreanini, R. Jackiw, Phys. Rev. D37, 2206 (1988).  A. Duncan, H.
Meyer-Ortmanns, R. Roskies, Phys. Rev. D36, 3788 (1987).

\bibitem{ai} A. Ashtekar, C.J. Isham, Class. and Quant. Grav. 9, 1069
(1992).  A. Ashtekar, J Lewandowski, in "Knots and Quantum Gravity",
ed J Baez, Oxford University Press (1993).

\bibitem{hamc} MP Blencowe, Nucl. Phys. B341, 213 (1990). V.
Husain, Hamiltonian constraint of quantum
general relativity,  Nucl. Phys. B313, 711-724 (1988).
 B. Br\"{u}gmann and J. Pullin, Nucl. Phys.
B363, 221 (1991).

\bibitem{brug} B Br\"ugmann, {\it The Hamiltonian Constraint in the
Loop Representation.} Ph.D. Thesis 1992. Unpublished.


\end{thebibliography}
\end{document}